\documentclass[review,number]{elsarticle}
\biboptions{sort&compress}

\usepackage{amssymb}
\usepackage{amsmath}

\journal{Aerospace Science and Technology}

\usepackage{amsmath,amssymb}
\usepackage{graphicx}
\usepackage{booktabs}
\usepackage{bm}
\usepackage{array}
\usepackage{placeins}
\usepackage{xurl}
\usepackage{xcolor}
\newcommand{\liwei}[1]{#1}

\journal{Aerospace Science and Technology}

\begin{document}

\begin{frontmatter}

\title{Learning Stiffness-Dependent Fluid–Structure Dynamics from Coarse Flow Representations}
\author[1]{Chun-Jun Pu}
\author[2]{Li-Wei Chen\corref{cor1}}
\ead{liwei.chen.aero@gmail.com}
\author[1]{Hai-Bo Huang\corref{cor1}}
\ead{huanghb@ustc.edu.cn}

\cortext[cor1]{Corresponding author}
\address[1]{University of Science and Technology of China, Hefei  230026, China}
\address[2]{Beijing Institute of Aeronautical Systems Engineering, Beijing 100076, China}
\begin{abstract}
\liwei{This paper develops a data-driven framework for long-term prediction of fluid--structure interaction (FSI) dynamics, focusing on the flow-induced vibration (FIV) of a flexible plate. A stiffness-conditioned neural evolution operator is developed by jointly representing the Eulerian flow field and the Lagrangian structural state. The flexible plate is described by 101 ordered structural tokens carrying nodal coordinates and velocities, while the nondimensional bending stiffness is introduced as a global conditioning variable.  Bidirectional cross-attention couples the fluid and structural representations within a hybrid CNN--Transformer architecture. Trained with staged multi-step autoregressive rollouts and symmetry-reflected trajectories, a single operator captures three distinct stiffness-dependent response regimes, including deflected--flapping, deflected, and flapping dynamics. The predicted trajectories preserve the principal flow organization, structural oscillation characteristics, and dominant frequencies, while blind 1000-step rollouts remain bounded. The same operator also interpolates to stiffness values excluded from training.}

To reduce the sensitivity of force reconstruction to under-resolved near-wall gradients, we construct a differentiable aerodynamic-force reconstruction module based on the derivative-moment transformation (DMT). By replacing conventional wall-stress surface integrals with an enclosed 2D curve integral surrounding the core vortex region, the aerodynamic lift and drag forces can be accurately reconstructed. This differentiable integration is implemented through a Signed Distance Function (SDF) and a smoothed Dirac-delta formulation, preserving gradient flow throughout the force-reconstruction procedure. Overall, the proposed framework provides an accurate and differentiable surrogate for stiffness-dependent FSI dynamics, offering a foundation for efficient parameter studies and future stiffness optimization in flow-energy-harvesting applications.
\end{abstract}


\end{frontmatter}


\section{Introduction}
\label{sec:intro}

Fluid--structure interaction (FSI) involving slender and flexible bodies arises in a wide range of engineering and biological systems, including aeroelastic structures, energy-harvesting devices, aquatic propulsion, and flow-responsive materials. These problems are governed by a nonlinear feedback loop: the fluid loading deforms the structure, while the resulting structural motion continuously modifies the fluid boundary and reorganizes the surrounding vortical flow. A representative configuration is the inverted flexible plate, whose downstream end is clamped while its upstream end is free. Depending on its nondimensional bending stiffness and other fluid--structural parameters, the plate can exhibit straight, deflected,
periodic flapping, and more complex oscillatory states \cite{kim2013flapping,sader2016large,goza2018global}. The transitions between these regimes are accompanied by substantial changes in the mean plate configuration, oscillation amplitude, dominant frequency, wake topology, and aerodynamic loading. These stiffness-dependent changes are also directly relevant to flow-energy-harvesting systems, whose performance depends strongly on the amplitude, frequency, and loading characteristics of the structural motion. Resolving such stiffness-dependent dynamics therefore requires a model that simultaneously represents the Eulerian flow field, the moving Lagrangian structure, and their strongly coupled temporal evolution.

High-fidelity numerical methods can reproduce these phenomena but remain computationally demanding when long trajectories or densely sampled parameter spaces are required. This cost becomes particularly restrictive in repeated-query applications such as parameter inference, uncertainty quantification, flow control, and aeroelastic design. Data-driven surrogate models provide a possible alternative by learning the evolution or solution operator directly from high-fidelity simulations. Early operator-learning frameworks, represented by DeepONet and the Fourier neural operator (FNO), demonstrated that mappings between infinite-dimensional function spaces can be approximated without solving each new PDE instance from the beginning \cite{lu2021learning,li2021fourier}. Graph-based simulators further extended learned physical modeling to irregular meshes and interacting discretized objects \cite{pfaff2021learning}. More recently, attention-based neural operators, including OFormer and GNOT, have provided flexible
representations of nonlocal spatial interactions, irregular sampling, multiple input functions, and multiscale PDE solutions \cite{li2023transformer,pmlr-v202-hao23c}.

Applying these ideas to FSI is more difficult than predicting a fluid field
over a fixed domain. The structural geometry is itself part of the evolving
physical state and determines where momentum is exchanged between the fluid
and the solid. A fluid-only Eulerian surrogate must infer this moving
interface implicitly from the velocity and pressure fields, so small
interface errors can be reintroduced at every autoregressive step. Existing
studies have explored CNN-based moving-boundary prediction, graph
representations of coupled systems, and FNO-based hybrid FSI solvers
\cite{bublik2023neural,Gao_2024,Xiao_2024}. Differentiable hybrid approaches
have also embedded coarse numerical FSI models within trainable recurrent
architectures, allowing gradients to propagate through the coupled rollout
trajectory \cite{fan2024differentiable}. In parallel, codomain-attention
operators have shown that tokenizing and exchanging information among
different physical variables can improve the representation of
multiphysics systems \cite{NEURIPS2024_bc75fa98}. Nevertheless, explicit
learned coupling between a dense Eulerian flow representation and an
ordered Lagrangian flexible structure remains comparatively underexplored.
In particular, directly concatenating a geometry mask with the fluid
channels does not preserve the distinct physical identities and
discretizations of the fluid and structural states.

A second challenge concerns long-horizon prediction. Neural PDE models are
often optimized using one-step or short-window errors, whereas practical FSI
applications require repeated reuse of predicted states. Even a small one-step
bias can accumulate during autoregressive rollout, leading to phase drift,
blurred vortical structures, duplicated moving interfaces, or collapse toward
an unphysical steady state. Multi-step message-passing solvers,
refinement-based models, and denoising operator Transformers have been
proposed to improve the stability of long PDE trajectories
\cite{brandstetter2022message,NEURIPS2023_d529b943,pmlr-v235-hao24d}.
Recent systematic comparisons of one-step, differentiably unrolled, and
gradient-truncated unrolled training further show that temporal unrolling
improves autoregressive inference by exposing the model to its own rollout
distribution. Full temporal differentiability provides additional benefits
within matched training configurations, while non-differentiable unrolling can
still retain much of the improvement associated with reducing the rollout
distribution shift \cite{list2025differentiability}. Benchmark studies have
further shown that autoregressive performance depends strongly on the training
horizon, noise treatment, temporal resolution, and underlying PDE dynamics
\cite{NEURIPS2024_d9875ebc}.

For FSI, this difficulty is compounded by the need to preserve not only
pointwise fields but also the coupled oscillatory attractor, including the
mean deformation, dominant frequency, phase-space structure, and wake
organization. Moreover, most existing surrogate models are demonstrated
for a single physical condition or a limited response regime. Whether one
conditional evolution operator can reproduce multiple
stiffness-dependent FSI regimes while remaining stable under a fully
autoregressive rollout is still insufficiently established.

A third challenge is the recovery of aerodynamic forces from
neural-predicted flow fields. Lift and drag are global functionals of the
coupled solution and provide a more integrated physical assessment than
pointwise field errors. Conventional force evaluation uses pressure and
viscous tractions on the instantaneous solid boundary. On a coarse neural
grid, however, the thin boundary layer and near-wall gradients are
generally under-resolved, making direct surface-stress integration
sensitive to local smoothing and differentiation errors.

Control-surface formulations offer an alternative by transferring force
evaluation from the body surface to a better-resolved outer-flow region.
\liwei{The} derivative-moment transformation (DMT) expresses the unsteady force
through vorticity moments and flux terms defined over a control domain and
its enclosing contour \cite{wu2005unsteady,WU_2007}. Control-surface force
measurements and more recent weighted-integral developments have
demonstrated the accuracy and diagnostic value of this class of
formulations \cite{Lentink2018,Gao_2025}.

For integration with neural models, force accuracy alone is not sufficient:
the force-reconstruction operator should also retain derivatives with
respect to the predicted velocity and pressure fields. Differentiable
surrogate-based design has demonstrated the value of combining level-set
geometry descriptions, automatic differentiation, and neural flow
predictions \cite{chen2021numerical}. Fully differentiable and GPU-accelerated CFD
platforms have further enabled gradient-based inverse modeling and
optimization through complete turbulence and FSI simulations
\cite{fan2026diffflowfsi}. Such approaches, however, generally require
differentiating through the numerical solver itself. A complementary
capability is to construct a differentiable force readout directly on
coarse flow fields produced by a learned evolution model. This requires a
continuous grid-compatible representation of the control contour and a
regularized integration rule that preserves both the DMT force identity and
a usable gradient path.

In this work, we develop a stiffness-conditioned neural evolution framework
for long-horizon prediction and force-based assessment of a two-dimensional
inverted flexible plate at $Re=200$. The model evolves the fluid velocity and
pressure on a fixed Eulerian grid while independently tracking the nodal
coordinates and velocities of 101 ordered Lagrangian plate points. 
\liwei{A convolutional encoder extracts spatial features from the flow field,
which are then organized into a compact sequence of fluid tokens for
Transformer attention.}
The position and velocity of each plate point are jointly embedded as a structural
token, while the nondimensional bending stiffness is introduced through an
additional global conditioning token. Bidirectional cross-attention transfers
information from the fluid to the structure and from the updated structure
back to the fluid.

The training data comprise 36 original stiffness-dependent trajectories,
supplemented by eight reflected trajectories. To improve recursive prediction,
the model is first trained using four-step rollouts and is subsequently
fine-tuned using eight-step rollouts. Starting from a single coupled state,
the complete fluid--structure state is advanced recursively without
intermediate ground-truth correction. The resulting model provides a unified
description of three stiffness-dependent response regimes, with particular
attention to the deflected and flapping dynamics.

\liwei{The principal contributions of this work are summarized as follows.}

\noindent\textbf{(1) Coupled Eulerian--Lagrangian representation for flexible-plate FSI.}
A neural evolution operator is developed to jointly represent the
Eulerian flow field and the Lagrangian flexible structure, with
bidirectional cross-attention coupling the two physical states. The
structural state includes both nodal coordinates and velocities, while
the nondimensional bending stiffness is introduced as a global
conditioning variable.

\noindent\textbf{(2) Unified autoregressive prediction across multiple
stiffness-dependent FSI regimes.}
A single stiffness-conditioned operator is trained over 36 bending
stiffnesses and captures the deflected--flapping, deflected, and flapping
regimes in a fully autoregressive manner. The predicted dynamics remain
stable over long rollouts and interpolate to unseen stiffness conditions.

\noindent\textbf{(3) Differentiable aerodynamic-force reconstruction from
coarse neural flow fields.}
A differentiable force-reconstruction operator is developed by combining
the DMT with a fixed control contour represented by a signed distance
function and a smoothed Dirac-delta formulation. The resulting
formulation recovers aerodynamic forces from coarse predicted flow
fields while retaining gradients with respect to the flow variables.

The remainder of this paper is organized as follows. Section~2 describes the
physical FSI system, numerical solver, multi-stiffness dataset, data
augmentation, and autoregressive sequence construction. Section~3 presents
the coupled CNN--Transformer evolution operator, the staged rollout-training
strategy, and the differentiable DMT force-reconstruction method. Section~4
evaluates the prediction accuracy across the stiffness parameter space,
representative multi-regime dynamics, aerodynamic-force reconstruction,
long-horizon rollout behavior, and interpolation to unseen stiffnesses.
The main conclusions and limitations are summarized in Section~5.

\section{Problem Description and FSI System}
\label{sec:problem}
We consider a two-dimensional flexible plate immersed in an incompressible viscous fluid flow. The structural boundary of the flexible plate is discretized into a set of $N_s = 101$ Lagrangian points. Each Lagrangian point $i \in \{1, \dots, N_s\}$ is characterized by its instantaneous Cartesian coordinates $\mathbf{x}_i(t) = [x_i(t), y_i(t)]^T$ and structural velocity $[u_{s,i}(t),v_{s,i}(t)]^{T}$.  A single nondimensional bending stiffness $\kappa$ is assigned uniformly to the plate in each simulation. The fluid state is defined on an Eulerian grid covering the computational domain $\Omega$and consists of the velocity $\boldsymbol{u}=[u,v]^{T}$ and pressure$p$, while the vorticity $\omega$ is derived from the velocity field. The coupling at the fluid-solid interface dictates that the fluid exerts aerodynamic forces on the plate, causing large-amplitude non-linear oscillations, which in turn modulate the vortex shedding patterns in the wake, exhibiting multiple distinct limit-cycle oscillation (LCO) modes as the nondimensional bending stiffness varies.

\begin{figure*}[t]
    \centering
    \includegraphics[width=\textwidth]
    {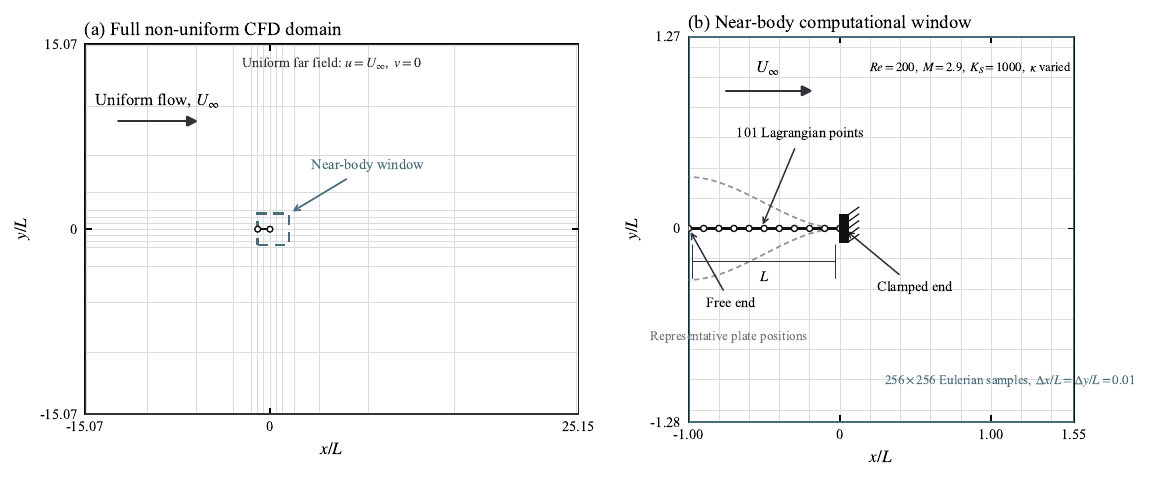}
    \caption{
    Physical configuration and computational domains of the flexible-plate
    FSI problem.
    (a) Full non-uniform CFD domain and the location of the near-body
    prediction window.
    (b) Near-body Eulerian window used by the neural evolution operator.
    }
    \label{fig:physical_domain}
\end{figure*}

The physical configuration and the computational domains are illustrated
in Fig.~\ref{fig:physical_domain}. The flexible plate is aligned with the
incoming uniform flow in its undeformed configuration, with its downstream
end clamped and its upstream end free. The full non-uniform CFD domain is
used to generate the reference trajectories, whereas a near-body window
containing the plate and the principal wake structures is extracted for
neural prediction.

\subsection{Physical configuration and governing equations}
\label{sec:physical_configuration}

The present study focuses on the flow-induced vibration of an inverted
flexible plate. The plate has a nondimensional length $L=1$ and is initially
aligned with the incoming flow.
Its downstream end is clamped at $(x,y)=(0,0)$, whereas its upstream end is
free. Thus, the undeformed plate occupies $-1\leq x\leq 0$ and $y=0$. A
uniform flow with velocity $\bm{u}_{\infty}=(U_{\infty},0)$ is imposed in the
positive $x$ direction. This inverted configuration permits large-amplitude
deformation of the upstream free end and produces a strongly coupled
interaction between the plate motion and the wake dynamics.

The fluid state is described by
\begin{equation}
    \bm{q}_{f}(\bm{x},t)
    =
    \left[u(\bm{x},t),v(\bm{x},t),p(\bm{x},t)\right]^{\mathrm T},
\end{equation}

where $u$ and $v$ are the Cartesian velocity components and $p$
denotes the pressure channel retained in the learning problem. Its
normalization relative to the pressure appearing in the governing
equations is specified below.
\begin{equation}
    \omega
    =
    \frac{\partial v}{\partial x}
    -
    \frac{\partial u}{\partial y}.
\end{equation}

The vorticity is not evolved as an independent variable by the neural model;
it is derived from the velocity field and is used to characterize the wake
regimes.

Using the plate length $L$, reference velocity $U_\infty$, fluid
density $\rho_f$, and convective time scale $L/U_\infty$, we define
the pressure appearing in the governing equations as
$p^\star=p_{\mathrm{dim}}/(\rho_f U_\infty^2)$.
The dimensionless incompressible Navier--Stokes equations are
\begin{align}
    \nabla\cdot\bm{u} &= 0, \\
    \frac{\partial\bm{u}}{\partial t}
    +\bm{u}\cdot\nabla\bm{u}
    &=-\nabla p^\star
    +\frac{1}{Re}\nabla^2\bm{u},
\end{align}
where
\begin{equation}
    Re=\frac{U_\infty L}{\nu}=200.
\end{equation}
and $\nu$ is the kinematic viscosity. The dimensionless fluid stress tensor is
written as
\begin{equation}
    \bm{\sigma}_f
    =-p^\star\bm{I}
    +\frac{1}{Re}
    \left(
    \nabla\bm{u}
    +\nabla\bm{u}^{T}
    \right).
\end{equation}

For the learning problem and the subsequent aerodynamic-force
reconstruction, the retained pressure channel $p$ is normalized by
the dynamic pressure $q_\infty=\tfrac{1}{2}\rho_f U_\infty^2$.
Consequently,
$p=p_{\mathrm{dim}}/q_\infty=2p^\star$.
This distinction is purely a normalization convention and does not
change the underlying dimensional pressure field.

The plate centreline is discretized by $N_s=101$ Lagrangian nodes connected
by 100 geometrically nonlinear beam elements. The structural state associated
with node $i$ is defined as
\begin{equation}
\boldsymbol{s}_i(t)
=
\left[
x_i(t),\,y_i(t),\,u_{s,i}(t),\,v_{s,i}(t)
\right]^{T},
\qquad
\boldsymbol{X}_s(t)
=
\left\{
\boldsymbol{s}_i(t)
\right\}_{i=1}^{N_s}
\in\mathbb{R}^{N_s\times4}.
\end{equation}

Although the structural solver retains the complete finite-element degrees of
freedom, only the in-plane nodal coordinates and velocities are retained in
the learning problem.

The principal control parameter is the nondimensional bending stiffness
\begin{equation}
    \kappa \equiv K_B
    =
    \frac{B}{\rho_f U_{\infty}^{2}L^{3}},
\end{equation}
where $B$ denotes the dimensional bending rigidity. The nondimensional axial
or extensional stiffness is held fixed at
\begin{equation}
    K_S=1000.
\end{equation}

Although the formulation permits a spatially varying local stiffness $k_i$,
all Lagrangian points in each simulation are assigned the same value, such
that $k_i=\kappa$ for $i=1,\ldots,N_s$.
The nondimensional mass ratio and Poisson ratio are fixed at
\begin{equation}
    m^{\ast}
    =
    \frac{\rho_s h}{\rho_f L}
    =2.9,
    \qquad
    \nu_s=0.25,
\end{equation}

respectively, while the structural damping coefficients are set to zero. The
fluid traction is transferred to the Lagrangian structure, and the no-slip
condition requires the interpolated fluid velocity to match the local
structural velocity on the moving plate. Consequently, the plate deformation
modulates the vortex formation and shedding process, while the resulting
fluid loading drives the structural motion.

\subsection{Coupled numerical solver}
\label{sec:numerical_solver}

The high-fidelity trajectories are generated using a partitioned FSI solver.
The incompressible flow is advanced by a D2Q9 multiple-relaxation-time lattice
Boltzmann method on a nonuniform Cartesian grid. The full fluid grid contains
$1468\times985$ points and covers
\begin{equation}
    -15.073\leq x\leq25.150,
    \qquad
    -15.073\leq y\leq15.073.
\end{equation}
The mesh is refined in the vicinity of the plate and gradually stretched
towards the far field. A uniform far-field state is imposed at the outer
boundaries.

The fluid and structure are coupled through an immersed-boundary forcing
procedure. At every coupled time step, the fluid variables are first advanced,
the immersed-boundary force is evaluated and transferred to the plate, and
the structural state is subsequently updated. The plate dynamics are solved
using a geometrically nonlinear finite-element formulation with a Newmark
time-integration scheme. The fluid reference values used by the source solver
are $L=1$, $U_{\infty}=0.025$, and $\rho_f=1$, yielding a convective reference
time $T_{\mathrm{ref}}=L/U_{\infty}=40$. The internal solver time step is
$\Delta t=0.01$, corresponding to the dimensionless interval
\begin{equation}
    \frac{\Delta t}{T_{\mathrm{ref}}}=2.5\times10^{-4}.
\end{equation}

The flow and structural states used for data construction are written every
$0.1$ dimensionless time units. Before storage, the nodal coordinates are scaled by $L$, both the
fluid and structural velocity components are scaled by $U_\infty$,
and the retained pressure channel $p$ is scaled by the dynamic pressure
$q_\infty=\tfrac{1}{2}\rho_f U_\infty^2$. Thus, the stored pressure
satisfies $p=2p^\star$. Dimensional quantities are denoted explicitly
only when needed hereafter.

\subsection{Multi-stiffness FSI dataset}
\label{sec:fsi_dataset}

Independent simulations are conducted by varying only the nondimensional
bending-stiffness parameter $\kappa$, while all other fluid and structural
parameters remain unchanged. The reference dataset consists of 36 independently
simulated flexible-plate cases, with $\kappa$ uniformly sampled from 0.05 to
0.40 at an interval of 0.01. The set of independently simulated stiffnesses is
therefore defined as
\begin{equation}
    \mathcal{K}_{\mathrm{train}}
    =
    \left\{
        0.05 + 0.01j
        \,\middle|\,
        j=0,1,\ldots,35
    \right\}.
    \label{eq:training_stiffness_set}
\end{equation}
Collectively, these 36 stiffness conditions represent the characteristic plate vibration modes and coupled response regimes considered in this study. The reflection-based augmentation used to construct the final training set is described in Section~2.4.

\subsubsection{Stiffness-dependent response regimes}
\label{sec:response_regimes}

Previous studies have shown that the dynamical response of an isolated
inverted flexible plate can be classified into several characteristic
regimes, including the straight, flapping, deflected, and
deflected--flapping modes. In the present study,
the selected bending-stiffness range focuses on three dynamically active
regimes: the deflected--flapping, deflected, and flapping modes.

The structural response is characterized by the peak-to-peak
transverse displacement $A$ of the free tip, reported in normalized
form as
\begin{equation}
    \frac{A}{L}
    =
    \frac{
    \displaystyle\max_{t\in\mathcal{T}}y_{\mathrm{tip}}(t)
    -
    \displaystyle\min_{t\in\mathcal{T}}y_{\mathrm{tip}}(t)
    }{L},
\end{equation}
where $L$ is the undeformed plate length and $\mathcal{T}$ denotes
the statistically stationary sampling interval.

\begin{figure}
    \centering
    \includegraphics[width=\linewidth]
    {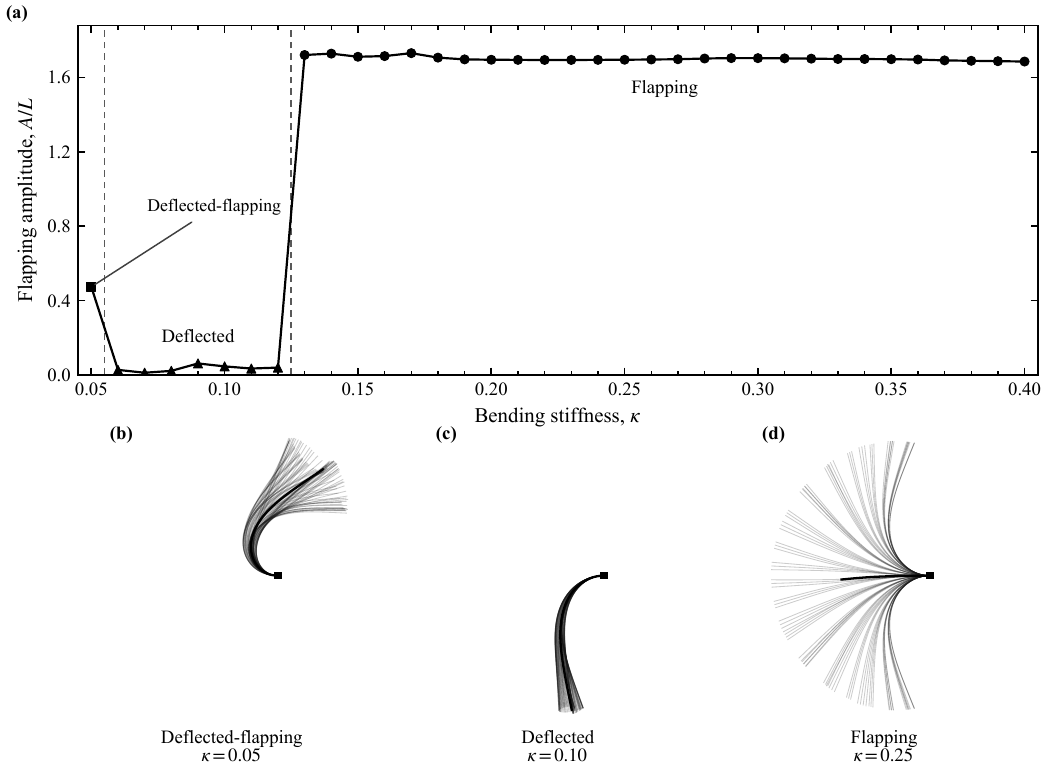}
    \caption{
    Stiffness-dependent response regimes of the isolated inverted flexible
    plate. (a) Normalized peak-to-peak transverse displacement of the free
    tip. (b--d) Representative full-body envelopes of the
    deflected--flapping, deflected, and flapping modes, respectively.
    The dashed lines visually separate the response regimes represented by
    the discrete stiffness samples.
    }
    \label{fig:stiffness_response}
\end{figure}

As shown in figure~\ref{fig:stiffness_response}, the densely sampled
stiffness cases reveal three distinct structural-response regimes. At
$\kappa=0.05$, the plate exhibits the deflected--flapping mode, characterized
by a strongly deflected mean configuration accompanied by finite-amplitude
oscillations. For $0.06\leq\kappa\leq0.12$, the response is classified as
the deflected mode, in which the plate oscillates weakly around a bent mean
configuration. When $0.13\leq\kappa\leq0.40$, the plate undergoes sustained
large-amplitude flapping across the undeformed centerline. These three
regimes collectively describe the characteristic stiffness-dependent
structural responses covered by the present dataset.

For each value of $\kappa$, a statistically developed segment over
\begin{equation}
    250.1\leq t\leq280.0
\end{equation}
is extracted from the original FSI simulation. The raw output interval is
$\Delta t_{\mathrm{raw}}=0.1$, giving 300 synchronized fluid--structure
snapshots for each condition. This late-time interval is selected to focus on
the developed stiffness-dependent oscillatory responses rather than the
initial transient from the undeformed state. For every stiffness condition,
the selected trajectory contains at least two complete oscillation cycles,
thereby retaining the essential periodic evolution of both the plate motion
and the associated wake dynamics.

For neural prediction, a local region surrounding the moving plate and its
near wake is extracted from the full numerical domain. The learning domain is
\begin{equation}
    \Omega_{\mathrm{ML}}
    =
    [-1.00,1.55]\times[-1.28,1.27]
\end{equation}
and is represented on a uniform $256\times256$ Cartesian grid. Each stored CFD snapshot contains the velocity components $u$ and $v$, the pressure $p$, and a binary geometry mask. Since the moving plate is represented independently by its Lagrangian state, the geometry mask is not supplied to or predicted by the neural evolution operator. The fluid state used in the learning problem is therefore
\begin{equation}
\boldsymbol{Q}_f^n
=
\left[
u^n,\,v^n,\,p^n
\right]
\in
\mathbb{R}^{3\times256\times256}.
\end{equation}
The corresponding structural state contains the nodal positions and
velocities,
\begin{equation}
\boldsymbol{X}_s^n
=
\left[
x_i^n,\,y_i^n,\,u_{s,i}^n,\,v_{s,i}^n
\right]_{i=1}^{101}
\in
\mathbb{R}^{101\times4}.
\end{equation}

To reduce temporal redundancy, one snapshot is retained from every five raw
outputs. The effective time interval of the learning problem is therefore
\begin{equation}
    \Delta t_{\mathrm{ML}}
    =5\Delta t_{\mathrm{raw}}
    =0.5.
\end{equation}
Each stiffness condition consequently contains 60 frames at
\begin{equation}
    t=250.1,250.6,\ldots,279.6.
\end{equation}
The 36 original stiffness conditions contain 2160 synchronized
fluid--structure states. Eight additional reflected trajectories, each
containing 60 frames, contribute another 480 states. The augmented training
set therefore contains 44 trajectories and 2640 synchronized
fluid--structure states in total.

\begin{table}[t]
    \centering
    \caption{Summary of the physical system and multi-stiffness dataset.}
    \label{tab:dataset_summary}
    \begin{tabular}{>{\raggedright\arraybackslash}p{0.52\linewidth}c}
        \toprule
        Quantity & Value \\
        \midrule
        Reynolds number & $Re=200$ \\
        Plate length & $L=1$ \\
        Mass ratio & $m^{\ast}=2.9$ \\
        Poisson ratio & $\nu_s=0.25$ \\
        Fixed extensional stiffness & $K_S=1000$ \\
        Training bending stiffnesses & $0.05$--$0.40$, $\Delta\kappa=0.01$ (36 cases) \\
        Structural discretization & 101 nodes, 100 beam elements \\
        Full fluid grid & $1468\times985$ \\
        Learning grid & $256\times256$ \\
        Learning-domain extent & $[-1.00,1.55]\times[-1.28,1.27]$ \\
        Raw snapshots per stiffness & 300 \\
        Raw output interval & $\Delta t_{\mathrm{raw}}=0.1$ \\
        Temporal subsampling factor & 5 \\
        Learning interval & $\Delta t_{\mathrm{ML}}=0.5$ \\
        Subsampled frames per stiffness & 60 \\
        Predicted fluid variables & $u,v,p$ \\
        Predicted structural variables & $x_i,y_i,u_{s,i},v_{s,i}$, $i=1,\ldots,101$ \\
        \bottomrule
    \end{tabular}
\end{table}

\subsection{Preprocessing, reflection augmentation, and autoregressive
sequence construction}
\label{sec:dataset_preprocessing}

To reduce the directional imbalance associated with the deflected responses,
reflection-based data augmentation is applied to the trajectories at
\[
\mathcal{K}_{\mathrm{mirror}}
=
\{0.05,0.06,\ldots,0.12\}.
\]

For a reflection about the plate centreline, the fluid variables are
transformed according to
\begin{equation}
u^{\mathrm{mir}}(x,y,t)=u(x,-y,t),\qquad
v^{\mathrm{mir}}(x,y,t)=-v(x,-y,t),\qquad
p^{\mathrm{mir}}(x,y,t)=p(x,-y,t),
\end{equation}

while the structural state is transformed as
\begin{equation}
x_i^{\mathrm{mir}}=x_i,\qquad
y_i^{\mathrm{mir}}=-y_i,\qquad
u_{s,i}^{\mathrm{mir}}=u_{s,i},\qquad
v_{s,i}^{\mathrm{mir}}=-v_{s,i}.
\end{equation}

The stiffness value is unchanged under reflection. This procedure introduces
eight additional trajectories without adding new stiffness values, resulting
in 36 original trajectories and eight reflected trajectories for training.

The velocity and pressure channels have different numerical scales. Global minimum and maximum values are evaluated separately for $q_c\in\{u,v,p\}$ over all 44 original and reflected training trajectories. The fluid channels are then normalized according to
\begin{equation}
    \widetilde q_c
    =
    \frac{q_c-q_{c,\min}}
    {q_{c,\max}-q_{c,\min}+10^{-8}}.
\end{equation}

The structural coordinates and velocities are retained in their nondimensional
physical scales without an additional dataset-wise min--max normalization.
The stiffness condition is mapped to $[-1,1]$ using
\begin{equation}
    \widetilde\kappa
    =
    2\frac{\kappa-\kappa_{\min}}
    {\kappa_{\max}-\kappa_{\min}}-1,
    \qquad
    \kappa_{\min}=0.05,
    \quad
    \kappa_{\max}=0.40.
\end{equation}

Training samples are constructed using multi-step autoregressive rollout
windows. Starting from an initial coupled state
$(\bm{Q}_f^n,\bm{X}_s^n)$, the model recursively predicts
\begin{equation}
    \left(
    \widehat{\bm{Q}}_f^{n+j},
    \widehat{\bm{X}}_s^{n+j}
    \right),
    \qquad j=1,\ldots,N_r,
\end{equation}

where $N_r$ denotes the rollout horizon adopted for a given training
configuration, and the prediction at step $j$ is used as the input to step
$j+1$. This construction allows the rollout horizon to be refined without
changing the definition of the underlying fluid--structure sequence. 

During training, zero-mean Gaussian perturbations with a standard deviation
of 0.005 are added at each rollout step to the normalized fluid variables and
all four components of the structural state. This exposes the network to off-trajectory states and improves its robustness during free autoregressive rollout.

The model adopted in this study is trained using a staged rollout strategy. It is first optimized with a four-step autoregressive horizon (\(N_r=4\)). The resulting R4 checkpoint is then used to initialize an eight-step autoregressive fine-tuning stage (\(N_r=8\)). The selected R4-to-R8 checkpoint is used for all neural predictions and aerodynamic-force reconstructions reported in this paper. For each 60-frame stiffness trajectory, the R4 and R8 stages contain 56 and 52 admissible rollout windows, respectively, corresponding to 2464 and 2288 training windows over the 44 original and reflected training trajectories.

The dataset spans stiffness-dependent changes in plate deformation,
oscillation amplitude, characteristic frequency, and wake topology. These
qualitative response regimes are first identified from the plate envelopes
and phase-aligned vorticity fields and are subsequently quantified using the
free-tip trajectory, dominant frequency, mean drag, and root-mean-square lift.
This organization provides the physical basis for evaluating whether a single
conditional neural evolution operator can reproduce multiple FSI regimes and
remain stable during long-term autoregressive prediction.

\section{Methodology}
\label{sec:methodology}

\begin{figure*}[t]
    \centering
    \includegraphics[width=\textwidth]
    {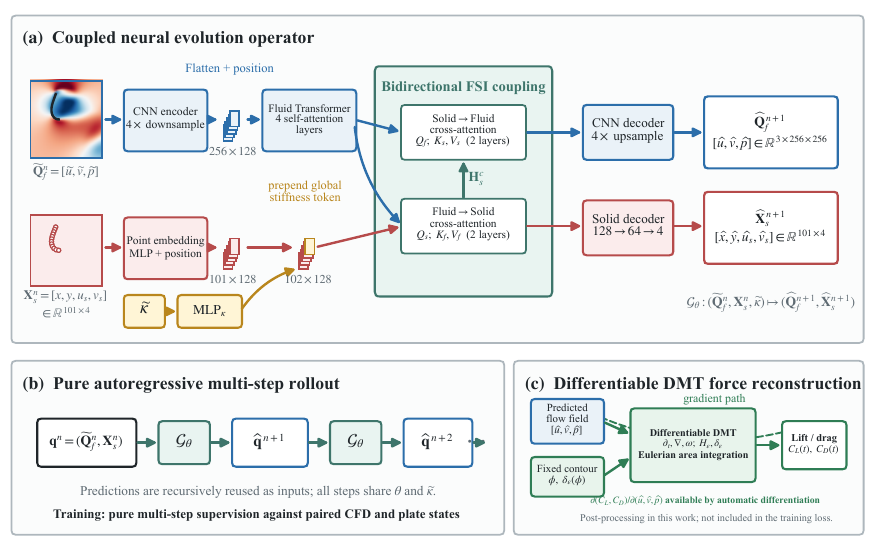}
    \caption{
    Overview of the proposed framework.
    (a) The Eulerian flow field and the Lagrangian plate states, including nodal positions and velocities, are encoded as fluid and structural tokens, respectively. The normalized
    bending stiffness is introduced as a global conditioning token, and
    bidirectional cross-attention exchanges information between the two
    physical representations.
    (b) The coupled state predicted at each time step is recursively reused
    as the input for subsequent autoregressive evolution.
    (c) Aerodynamic lift and drag are reconstructed from the predicted flow
    field using a differentiable DMT operator defined on a fixed
    level-set control contour. 
    }
    \label{fig:coupled_fsi_architecture}
\end{figure*}

The core of the proposed framework consists of two main pillars: a hybrid CNN-Transformer neural evolution operator that models the autoregressive spatiotemporal FSI coupling, and a differentiable aerodynamic force reconstruction module based on \liwei{the} DMT force formulation, facilitated by an SDF-based soft integration method. An overview of the complete framework is
shown in Fig.~\ref{fig:coupled_fsi_architecture}.

\subsection{CNN-Transformer Hybrid Evolution Operator}
To resolve the multi-scale characteristics of the flow field and the discrete nature of the structure, we map the Eulerian fluid field and the Lagrangian solid points into a unified latent space using a hybrid tokenization strategy.

At each discrete time step \(n\), the normalized fluid state is represented by
\begin{equation}
    \widetilde{\bm{Q}}_f^n
    =
    \left[
    \widetilde{u}^n,
    \widetilde{v}^n,
    \widetilde{p}^n
    \right]
    \in
    \mathbb{R}^{3\times H\times W},
    \qquad H=W=256.
\end{equation}
The instantaneous state of the flexible plate is represented by $N_s=101$
Lagrangian points,
\begin{equation}
\boldsymbol{X}_s^n
=
\left[
\boldsymbol{s}_1^n,\boldsymbol{s}_2^n,\ldots,
\boldsymbol{s}_{N_s}^n
\right]^{T}
\in
\mathbb{R}^{N_s\times4},
\qquad
\boldsymbol{s}_i^n
=
\left[
x_i^n,\,y_i^n,\,u_{s,i}^n,\,v_{s,i}^n
\right]^{T}.
\end{equation}
The dimensionless bending stiffness is supplied to the network as the
normalized scalar \(\widetilde{\kappa}\). The one-step coupled evolution
learned by the model is expressed as
\begin{equation}
    \left(
    \widehat{\widetilde{\bm{Q}}}_f^{\,n+1},
    \widehat{\bm{X}}_s^{\,n+1}
    \right)
    =
    \mathcal{G}_{\theta}
    \left(
    \widetilde{\bm{Q}}_f^n,
    \bm{X}_s^n,
    \widetilde{\kappa}
    \right),
    \label{eq:neural_evolution_operator}
\end{equation}

where \(\mathcal{G}_{\theta}\) denotes the neural evolution operator
parameterized by \(\theta\), and the hat denotes a predicted quantity.
Repeated application of equation~\eqref{eq:neural_evolution_operator}
produces an autoregressive rollout of the coupled fluid--structure dynamics.

\subsubsection{Fluid Tokenization via CNN}

At each time step \(n\), the normalized two-dimensional fluid state
\(\widetilde{\bm{Q}}_f^n\in\mathbb{R}^{3\times H\times W}\), consisting of
the velocity components \(\widetilde{u}\) and \(\widetilde{v}\) and the
pressure \(\widetilde{p}\), is processed by a convolutional encoder. The
purpose of the encoder is to extract hierarchical spatial features and
convert the dense Eulerian flow field into a compact sequence of fluid
tokens suitable for global attention.

The convolutional encoder consists of four successive downsampling blocks.
Each block employs a \(4\times4\) convolution with stride 2 and padding 1,
followed by batch normalization and a GELU activation. The spatial resolution
and channel dimension evolve as
\begin{equation}
\begin{aligned}
3\times256\times256
&\rightarrow32\times128\times128
\rightarrow64\times64\times64\\
&\rightarrow128\times32\times32
\rightarrow128\times16\times16.
\end{aligned}
\label{eq:fluid_cnn_dimensions}
\end{equation}
The resulting feature map can be expressed as
\begin{equation}
    \bm{F}_f^n
    =
    \mathcal{E}_{\mathrm{CNN}}
    \left(\widetilde{\bm{Q}}_f^n\right)
    \in
    \mathbb{R}^{d_m\times16\times16},
    \qquad d_m=128,
\label{eq:fluid_cnn_encoding}
\end{equation}
where \(\mathcal{E}_{\mathrm{CNN}}\) denotes the convolutional encoder and
\(d_m\) is the latent embedding dimension.

The feature map is subsequently flattened along its two spatial dimensions
to produce \(N_f=16\times16=256\) fluid tokens. A learnable positional
embedding \(\bm{P}_f\) is added to retain the spatial identity of each token:
\begin{equation}
    \bm{H}_{f,0}^n
    =
    \operatorname{Flatten}
    \left(\bm{F}_f^n\right)
    +
    \bm{P}_f
    \in
    \mathbb{R}^{N_f\times d_m},
    \qquad N_f=256.
\label{eq:fluid_tokens}
\end{equation}
Before interacting with the structural representation, the fluid-token
sequence is processed by a four-layer Transformer encoder. Each layer
contains four-head self-attention, a pre-normalization structure, and a
feed-forward network with hidden dimension \(4d_m=512\). The encoded fluid
representation is therefore given by
\begin{equation}
    \bm{H}_f^n
    =
    \mathcal{T}_f
    \left(\bm{H}_{f,0}^n\right)
    \in
    \mathbb{R}^{N_f\times d_m},
\label{eq:fluid_transformer}
\end{equation}
where \(\mathcal{T}_f\) denotes the fluid Transformer encoder.

This hybrid representation is particularly suitable for the present FSI
problem. The convolutional encoder extracts local spatial features associated
with velocity gradients and vortical structures while reducing the original
\(256\times256\) Eulerian grid to only 256 latent tokens. Applying
self-attention directly to all \(256^2\) grid points would be computationally
prohibitive because its cost increases quadratically with the sequence
length. The CNN-based tokenization reduces the attention sequence length by
a factor of 256, making global interaction modeling computationally
tractable.

The subsequent Transformer encoder complements the local
receptive fields of the convolutional layers by capturing nonlocal
dependencies between the near-plate flow and the downstream wake. The
resulting fluid representation therefore combines local flow features,
long-range wake interactions, and computationally efficient global
attention before coupling with the flexible-plate representation.

\subsubsection{Solid Tokenization and Physical Embedding}
\label{sec:solid_tokenization}

The structural state is represented by $N_s=101$ ordered Lagrangian points.
Since this representation is already discrete, each point is directly
embedded as a structural token without further spatial downsampling. For the
$i$th point, the instantaneous nodal state
\[
\boldsymbol{s}_i^n
=
\left[
x_i^n,\,y_i^n,\,u_{s,i}^n,\,v_{s,i}^n
\right]^{T}
\]
is mapped into the latent space by an embedding MLP and combined with a
learnable positional embedding:
\begin{equation}
\boldsymbol{h}_{s,i}^{n,0}
=
\operatorname{MLP}_s\left(\boldsymbol{s}_i^n\right)
+
\boldsymbol{p}_{s,i},
\qquad
i=1,\ldots,N_s.
\end{equation}
Here, $\boldsymbol{p}_{s,i}$ distinguishes the fixed ordering of the
Lagrangian points, while $\boldsymbol{s}_i^n$ describes both the instantaneous
geometry and kinematics of the plate. The resulting structural-token sequence
is
\begin{equation}
\boldsymbol{H}_{s,0}^{n}
=
\left[
\boldsymbol{h}_{s,1}^{n,0},
\ldots,
\boldsymbol{h}_{s,N_s}^{n,0}
\right]^{T}
\in
\mathbb{R}^{N_s\times d_m},
\qquad
d_m=128.
\end{equation}

The normalized bending stiffness \(\widetilde{\kappa}\) is represented by
an additional token,
\begin{equation}
    \bm{h}_{\kappa}^{0}
    =
    \operatorname{MLP}_{\kappa}
    \left(\widetilde{\kappa}\right)
    +
    \bm{p}_{\kappa},
\label{eq:kappa_token}
\end{equation}
which is prepended to the structural-token sequence:
\begin{equation}
    \bm{H}_{s\kappa,0}^n
    =
    \left[
    \bm{h}_{\kappa}^{0};
    \bm{H}_{s,0}^n
    \right]
    \in
    \mathbb{R}^{(N_s+1)\times d_m}
    =
    \mathbb{R}^{102\times128}.
\label{eq:solid_kappa_tokens}
\end{equation}
The stiffness token provides a global conditioning variable for the
structural representation and participates together with the Lagrangian
tokens in the subsequent cross-attention operations.

\subsubsection{Bidirectional Cross-Attention Coupling Mechanism}
The fundamental FSI interaction is modeled via a bidirectional cross-attention mechanism, alternating between Fluid-to-Solid ($f \to s$) and Solid-to-Fluid ($s \to f$) blocks.
For the $f \to s$ coupling:
\begin{equation}
\widehat{\mathbf{H}}_{s}^{n}
=
\operatorname{softmax}
\left(
\frac{\mathbf{Q}_{s}\mathbf{K}_{f}^{T}}{\sqrt{d_k}}
\right)
\mathbf{V}_{f}
+
\mathbf{H}_{s}^{n}.
\end{equation}

where $\mathbf{Q}_{s}$ comes from the solid tokens, while
$\mathbf{K}_{f}$ and $\mathbf{V}_{f}$ come from the fluid tokens.
\begin{equation}
\widehat{\mathbf{H}}_{f}^{n}
=
\operatorname{softmax}
\left(
\frac{\mathbf{Q}_{f}\mathbf{K}_{s}^{T}}{\sqrt{d_k}}
\right)
\mathbf{V}_{s}
+
\mathbf{H}_{f}^{n}.
\end{equation}

Both directional coupling modules are implemented using two-layer
Transformer decoders. Each decoder layer consists of four-head attention,
pre-normalization, residual connections, and a feed-forward network with
hidden dimension \(4d_m=512\). For \(d_m=128\), the feature dimension of
each attention head is \(d_k=32\).

The fluid-to-solid attention operates between \(N_s+1=102\) structural
tokens, including the stiffness token, and \(N_f=256\) fluid tokens. Its
attention matrix therefore has the dimension
\begin{equation}
    \bm A_{f\rightarrow s}
    \in
    \mathbb{R}^{(N_s+1)\times N_f}
    =
    \mathbb{R}^{102\times256}.
\end{equation}
The subsequent solid-to-fluid attention uses the updated structural
representation as its memory, giving
\begin{equation}
    \bm A_{s\rightarrow f}
    \in
    \mathbb{R}^{N_f\times(N_s+1)}
    =
    \mathbb{R}^{256\times102}.
\end{equation}
The two operations are evaluated sequentially, such that the structural
tokens are first conditioned on the fluid representation and the fluid
tokens are then updated using the coupled structural representation. This
formulation couples the Eulerian and Lagrangian representations in the
latent space without introducing an explicit pointwise interpolation
operator.

\subsubsection{Autoregressive Rollout and Pure Supervision}
The updated latent states are passed through the decoders to predict the next time step. The available trajectory for each stiffness contains $T_{\mathrm{seq}}=60$ frames. The network is trained autoregressively using pure multi-step supervision.

Following the sequence construction described in Section 2.4, a staged autoregressive training strategy is adopted. The network is first trained using a four-step rollout horizon (\(N_r=4\)). The resulting R4 checkpoint is subsequently used to initialize a second training stage with an eight-step rollout horizon (\(N_r=8\)). Within each training window, the predicted fluid and structural states are recursively fed back to the model without replacement by the corresponding reference states. All prediction and force-reconstruction results reported in this study are obtained using the selected R4-to-R8 checkpoint.

The training objective is defined as the sum of the fluid and structural mean-squared errors over the \(N_r\) recursively predicted steps. Thus, \(N_r=4\) in the initial training stage and \(N_r=8\) in the subsequent fine-tuning stage:

\begin{equation}
    \mathcal{L}
    =
    \mathcal{L}_{f}
    +
    \mathcal{L}_{s},
\label{eq:total_training_loss}
\end{equation}
where
\begin{equation}
\begin{aligned}
    \mathcal{L}_{f}
    &=
    \frac{1}{N_r}
    \sum_{j=1}^{N_r}
    \operatorname{MSE}
    \left(
    \widehat{\widetilde{\bm Q}}_f^{\,n+j},
    \widetilde{\bm Q}_f^{\,n+j}
    \right),\\
    \mathcal{L}_{s}
    &=
    \frac{1}{N_r}
    \sum_{j=1}^{N_r}
    \operatorname{MSE}
    \left(
    \widehat{\bm X}_s^{\,n+j},
    \bm X_s^{\,n+j}
    \right).
\end{aligned}
\label{eq:fluid_solid_losses}
\end{equation}

The structural loss is evaluated jointly over the two nodal-coordinate components and the two structural-velocity components.The two loss components are assigned equal weights in the present model.

\subsection{Differentiable Aerodynamic Force Reconstruction via DMT and SDF}

Standard aerodynamic-force evaluation relies on direct integration of
the pressure and viscous tractions over the moving solid boundary
$\partial B$. For neural-predicted flow fields on a relatively coarse
Eulerian grid, this evaluation can be sensitive to inaccuracies in the
near-wall pressure and velocity gradients. We therefore employ \liwei{the}
DMT formulation \cite{wu2005unsteady} on a fixed closed control contour
$\Sigma$ enclosing the plate in the better-resolved outer-flow region.

In its general vorticity-moment form, the total aerodynamic force
comprises the time derivative of the vorticity moment within the
enclosed domain $\Omega_c$ and a boundary flux integral along
$\Sigma$:
\begin{equation}
\mathbf{F} = -\rho \frac{d}{dt} \int_{\Omega_c} \mathbf{r} \times (\omega \mathbf{k}) d\Omega + \oint_{\Sigma} \mathbf{T}(\mathbf{u}, \omega, \mathbf{r}, \mathbf{n}) ds \label{eq:wu}
\end{equation}
where $\boldsymbol{r}$ is the position vector, $\boldsymbol{k}$ is the
out-of-plane unit vector, and $\boldsymbol{T}$ collectively represents
the pressure, viscous-vorticity, and momentum-flux contributions on the
control contour.

A critical challenge in implementing Eq.\ref{eq:wu} within an end-to-end neural network is rendering the curve integral differentiable with respect to the Eulerian grid representation. Explicitly tracking the curve $\Sigma$ breaks grid continuity. To achieve a fully differentiable line integral, we introduce a continuous Signed Distance Function (SDF) following the implicit boundary treatment methodology thoroughly established in deep learning fluid surrogates \cite{chen2021numerical}. Here, an SDF field $\phi(\mathbf{x})$ is constructed to implicitly represent the integration contour $\Sigma$:
\begin{equation}
\phi(\mathbf{x}) = 
\begin{cases} 
-d(\mathbf{x}, \Sigma) & \text{if } \mathbf{x} \in \Omega_c \\ 
0 & \text{if } \mathbf{x} \in \Sigma \\ 
d(\mathbf{x}, \Sigma) & \text{if } \mathbf{x} \in \Omega - \Omega_c 
\end{cases}
\end{equation}
where $d(\mathbf{x}, \Sigma)$ denotes the Euclidean distance from a point $\mathbf{x}$ to the curve $\Sigma$. The outward unit normal vector is rigorously obtained as $\mathbf{n} = \nabla \phi / \|\nabla \phi\|$.

We replace the hard line integral over $\Sigma$ with an area integral
over the Eulerian domain $\Omega$ using a regularized Heaviside
function $H_\epsilon$ and its derivative, the smoothed Dirac-delta
function $\delta_\epsilon(\phi)$. To remain consistent with the
discrete implementation, a compactly supported cosine regularization
is employed:
\begin{equation}
\begin{aligned}
H_\epsilon(\phi)
&=
\begin{cases}
0,
& \phi<-\epsilon, \\[2pt]
\displaystyle
\frac{1}{2}
\left[
1+\frac{\phi}{\epsilon}
+\frac{1}{\pi}
\sin\left(\frac{\pi\phi}{\epsilon}\right)
\right],
& |\phi|\leq\epsilon, \\[6pt]
1,
& \phi>\epsilon,
\end{cases}
\\[8pt]
\delta_\epsilon(\phi)
&=
\frac{\partial H_\epsilon}{\partial\phi}
=
\begin{cases}
\displaystyle
\frac{1}{2\epsilon}
\left[
1+\cos\left(\frac{\pi\phi}{\epsilon}\right)
\right],
& |\phi|\leq\epsilon, \\[6pt]
0,
& |\phi|>\epsilon.
\end{cases}
\end{aligned}
\end{equation}
where $\epsilon$ controls the half-width of the regularized narrow
band. For all aerodynamic-force results reported in this study,
$\epsilon=0.51\max(\Delta x,\Delta y)=0.0051$.

Consequently, a line integral of a generic integrand $I$ over the
curve $\Sigma$ is approximated by a differentiable two-dimensional
area integral weighted by the regularized Dirac-delta function:
\begin{equation}
    \oint_{\Sigma} I\,\mathrm{d}s
    \approx
    \int_{\Omega}
    I\delta_\epsilon(\phi)
    \lVert\nabla\phi\rVert\,\mathrm{d}\Omega .
\end{equation}
Applying this regularized SDF-based transformation to the general
\liwei{vorticity-moment}
equation gives the following
approximate area-integral representation:
\begin{equation}
\mathbf{F} \approx -\rho \frac{d}{dt} \int_{\Omega} \mathbf{r} \times (\omega \mathbf{k}) \, H_\epsilon(-\phi) \, d\Omega + \int_{\Omega} \mathbf{T} \, \delta_\epsilon(\phi) \, \|\nabla \phi\| \, d\Omega
\end{equation}

\liwei{The explicit-contour and SDF-based formulations retain the same physical
terms in the DMT force identity, differing only in how the
control-contour integrals are represented numerically.}

This formulation seamlessly maps the curve integral onto the 2D Eulerian grid predicted by the CNN. Since spatial integration on a fixed grid and point-wise multiplication with $\delta_\epsilon(\phi)$ are inherently tensor operations in modern deep learning frameworks, the entire force reconstruction process becomes rigorously differentiable. Error gradients from force-based loss functions can propagate back through the softened Dirac-delta weighting seamlessly to both the CNN fluid tokens and the Transformer's solid structural representation.

\section{Results and Discussion}
\label{sec:results}

\subsection{Prediction across stiffness-dependent response regimes}
\label{sec:regime-prediction}

\subsubsection{Representative regimes from one stiffness-conditioned operator}
\label{sec:representative-regimes}

Following the physical classification introduced in Section~2, three
stiffnesses are selected to represent the distinct responses contained in the
data set:
\begin{equation}
\kappa=0.05\quad\text{(deflected-flapping)},\qquad
\kappa=0.10\quad\text{(deflected)},\qquad
\kappa=0.30\quad\text{(flapping)}.
\end{equation}
All three trajectories are produced by the same trained operator; only the
stiffness-conditioning token and the initial coupled state are changed. The
first state is supplied from the corresponding CFD trajectory, after which
the fluid fields and the positions and velocities of the 101 plate nodes are
advanced recursively for 59 steps.

\begin{figure}[t]
    \centering
    \includegraphics[width=\textwidth]
    {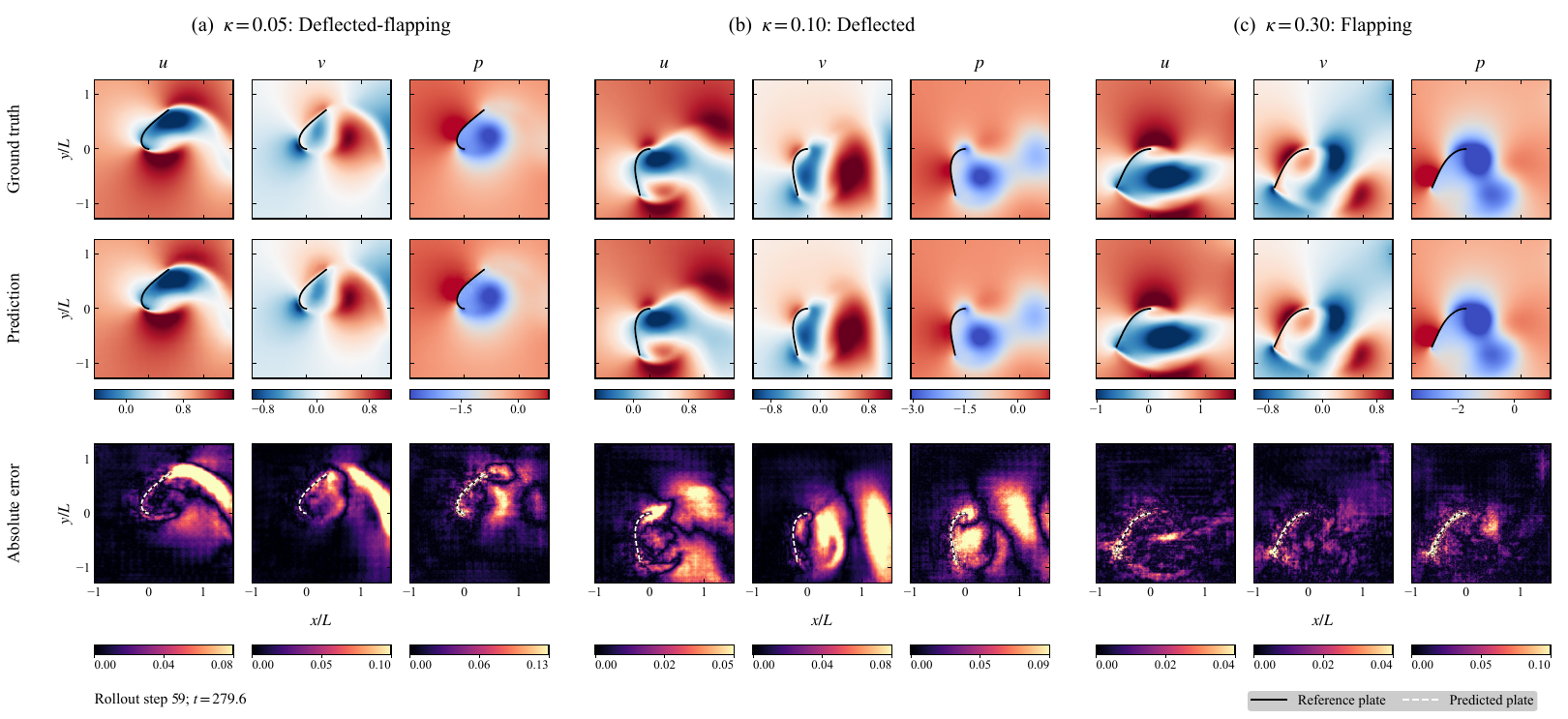}
    \caption{Flow-field reconstruction at rollout step 59 for three
    representative response regimes: (a) $\kappa=0.05$;
    (b) $\kappa=0.10$; and (c) $\kappa=0.30$. Within each group, the
    columns show $u$, $v$, and $p$, and the rows show the CFD ground truth,
    neural prediction, and pointwise absolute error.}
    \label{fig:representative-regime-fields}
\end{figure}

Figure~\ref{fig:representative-regime-fields} shows that the predicted plate
geometry and the principal high- and low-speed regions remain aligned with the
CFD solution after repeated autoregressive reuse. The distinction among the
three regimes is not imposed by a discrete class label. It emerges from
conditioning the common evolution map on $\kappa$ and from recursively
coupling the resulting structural motion back into the fluid representation.
In particular, the model retains the biased mean configuration at
$\kappa=0.05$, the small-amplitude deflected response at $\kappa=0.10$, and
the large-amplitude nearly symmetric flapping at $\kappa=0.30$.

The field and structural errors at rollout step $n$ are measured as
\begin{equation}
e_q^n
=
\frac{
\left\|
\widehat{q}^{\,n}-q^n
\right\|_2
}{
\left\|
q^n
\right\|_2
},
\qquad
q\in\{u,v,p\},
\label{eq:fluid-relative-error}
\end{equation}
and
\begin{equation}
e_s^n
=
\left[
\frac{1}{2N_s}
\sum_{i=1}^{N_s}
\left\|
\widehat{\boldsymbol{x}}_{s,i}^{\,n}
-
\boldsymbol{x}_{s,i}^{\,n}
\right\|_2^2
\right]^{1/2}.
\label{eq:solid-coordinate-rmse}
\end{equation}
Here,
$\bm{x}_{s,i}^{n}=(x_i^{n},y_i^{n})^{T}$ and
$\widehat{\bm{x}}_{s,i}^{n}
=(\widehat{x}_i^{n},\widehat{y}_i^{n})^{T}$
denote the reference and predicted positions of the $i$th Lagrangian
point, respectively. The quantity $e_q^n$ is the relative $L_2$ error
over the Eulerian grid, whereas $e_s^n$ is the coordinate RMSE over
the ordered Lagrangian points.

\begin{figure}[t]
    \centering
    \includegraphics[width=\textwidth]
    {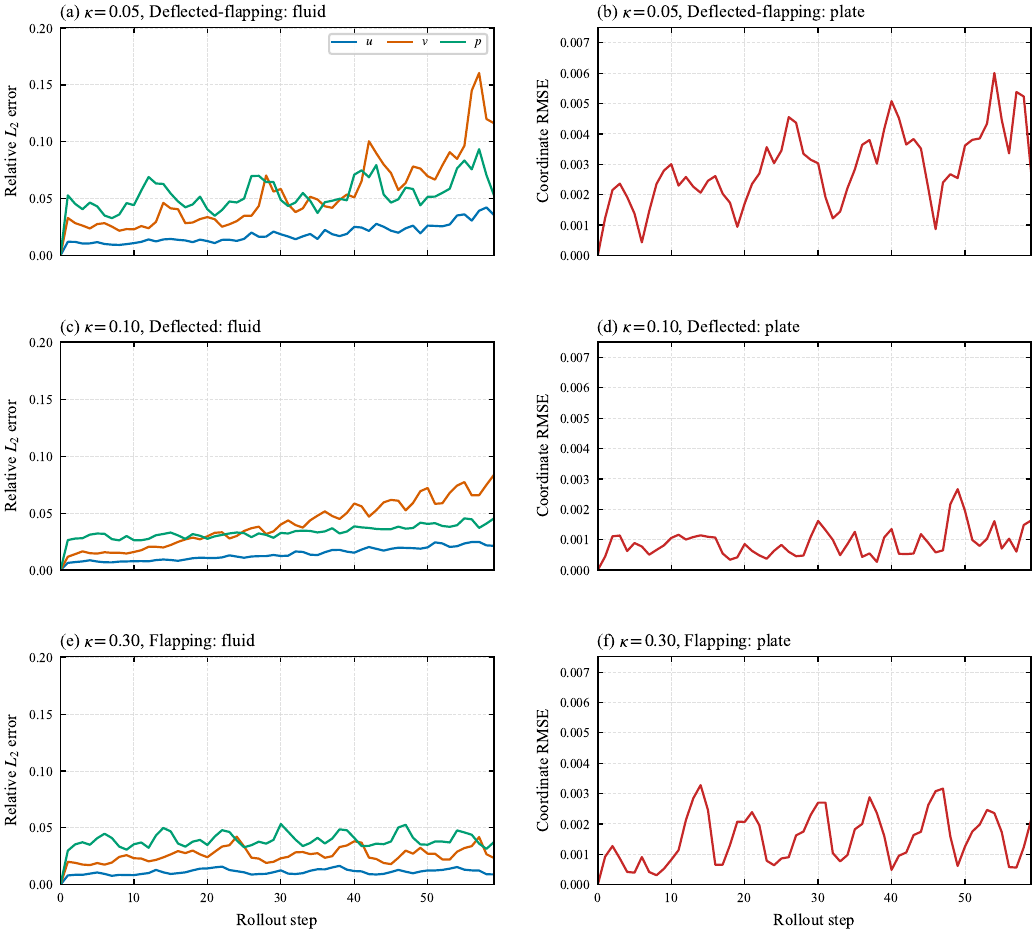}
    \caption{Stepwise rollout errors for the three representative response
    regimes over 59 predicted states. The left column shows the relative
    $L_2$ errors of $u$, $v$, and $p$, while the right column shows the
    coordinate RMSE of the plate. The rows correspond to
    $\kappa=0.05$, $0.10$, and $0.30$, respectively.}
    \label{fig:representative-rollout-errors}
\end{figure}

As shown in Fig.~\ref{fig:representative-rollout-errors}, the errors remain
bounded over the evaluated interval, although their magnitude depends on both
the variable and the response regime. Averaged over the 59 predicted states,
the relative errors are 1.89\%, 5.39\%, and 5.41\% for $u$, $v$, and $p$,
respectively, at $\kappa=0.05$; 1.41\%, 4.08\%, and 3.36\% at
$\kappa=0.10$; and 1.13\%, 2.63\%, and 3.94\% at $\kappa=0.30$.
The corresponding mean plate-coordinate RMSE values are
$2.91\times10^{-3}$, $9.17\times10^{-4}$, and
$1.53\times10^{-3}$, respectively. The plate-coordinate errors remain of the
same order across the three response regimes, indicating that the structural
evolution is preserved even in the large-amplitude flapping case.

\subsubsection{Accuracy across all 36 training stiffnesses}
\label{sec:all-stiffness-accuracy}

\begin{figure}[t]
    \centering
    \includegraphics[width=\textwidth]
    {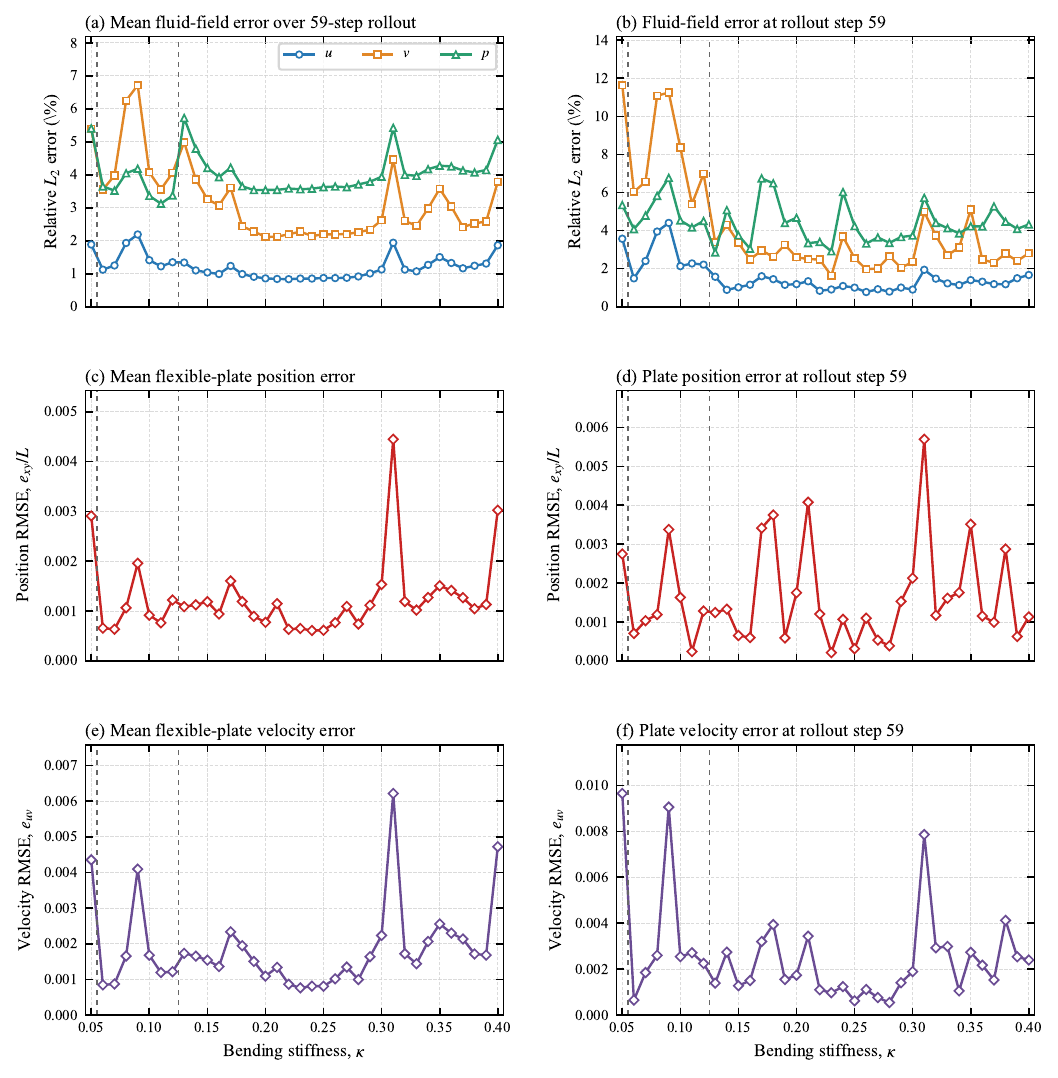}
    \caption{Prediction errors across the 36 training stiffnesses.
    (a) Mean fluid-field relative $L_2$ error over the 59 predicted states.
    (b) Fluid-field relative $L_2$ error at rollout step 59.
    (c) Mean flexible-plate position RMSE over the 59 predicted states.
    (d) Plate-position RMSE at rollout step 59.
    (e) Mean flexible-plate velocity RMSE over the 59 predicted states.
    (f) Plate-velocity RMSE at rollout step 59.
    The dashed separators indicate the deflected--flapping, deflected,
    and flapping regimes.}
    \label{fig:all-stiffness-errors}
\end{figure}

The representative cases provide a qualitative view of the three regimes,
whereas the complete stiffness sweep tests whether the same operator behaves
consistently throughout the parameter set. Figure~\ref{fig:all-stiffness-errors}
summarizes the mean and final-step errors for all 36 training stiffnesses.

Hereafter, the plate-coordinate RMSE defined in
Eq.~\eqref{eq:solid-coordinate-rmse} is denoted by $e_{xy}^{n}$. Since the
present model also predicts the two structural velocity components, the
corresponding plate-velocity RMSE is defined as
\begin{equation}
e_{uv}^{n}
=
\left[
\frac{1}{2N_s}
\sum_{i=1}^{N_s}
\left\|
\begin{bmatrix}
\widehat{u}_{s,i}^{\,n} \\
\widehat{v}_{s,i}^{\,n}
\end{bmatrix}
-
\begin{bmatrix}
u_{s,i}^{n} \\
v_{s,i}^{n}
\end{bmatrix}
\right\|_2^2
\right]^{1/2}.
\label{eq:solid-velocity-rmse}
\end{equation}

Across the 36 training conditions, the maximum mean relative errors are
approximately 2.19\% for $u$, 6.72\% for $v$, and 5.72\% for $p$.
The maxima for $u$ and $v$ occur at $\kappa=0.09$, whereas the maximum
mean pressure error occurs at $\kappa=0.13$. The maximum mean plate-position
and plate-velocity RMSE values are $4.45\times10^{-3}$ and
$6.22\times10^{-3}$, respectively, both occurring at $\kappa=0.31$.
Because the reported errors compare time-aligned snapshots point by point,
an accumulated phase difference between the predicted and reference
oscillations can produce a relatively large instantaneous error without
implying a qualitative failure of the predicted dynamics.

\begin{table}[t]
    \centering
    \caption{Mean errors over the 59 predicted states for eight representative
    stiffnesses selected from the 36 training conditions. Fluid errors are
    relative $L_2$ errors in percent; $e_{xy}$ and $e_{uv}$ are the
    nondimensional RMSEs of the plate position and velocity, respectively.}
    \label{tab:representative-stiffness-errors}
    \small
    \setlength{\tabcolsep}{4pt}
    \begin{tabular}{c l c c c c c}
        \hline
        $\kappa$ & Regime
        & $e_u$ (\%) & $e_v$ (\%) & $e_p$ (\%)
        & $e_{xy}$ & $e_{uv}$ \\
        \hline
        0.05 & Deflected--flapping
        & 1.89 & 5.39 & 5.41
        & $2.91\times10^{-3}$ & $4.35\times10^{-3}$ \\

        0.10 & Deflected
        & 1.41 & 4.08 & 3.36
        & $9.17\times10^{-4}$ & $1.68\times10^{-3}$ \\

        0.15 & Flapping
        & 1.04 & 3.26 & 4.21
        & $1.19\times10^{-3}$ & $1.54\times10^{-3}$ \\

        0.20 & Flapping
        & 0.86 & 2.12 & 3.54
        & $7.75\times10^{-4}$ & $1.09\times10^{-3}$ \\

        0.25 & Flapping
        & 0.87 & 2.20 & 3.63
        & $6.15\times10^{-4}$ & $8.10\times10^{-4}$ \\

        0.30 & Flapping
        & 1.13 & 2.63 & 3.94
        & $1.53\times10^{-3}$ & $2.23\times10^{-3}$ \\

        0.35 & Flapping
        & 1.51 & 3.58 & 4.28
        & $1.51\times10^{-3}$ & $2.56\times10^{-3}$ \\

        0.40 & Flapping
        & 1.86 & 3.79 & 5.06
        & $3.02\times10^{-3}$ & $4.72\times10^{-3}$ \\
        \hline
    \end{tabular}
\end{table}

\FloatBarrier

Despite these localized error variations, the neural operator reproduces
the overall oscillatory evolution of the coupled system and preserves the
principal spatial organization of the velocity and pressure fields over the
evaluated rollout. More importantly, the same stiffness-conditioned operator
successfully captures the deflected--flapping, deflected, and flapping
response regimes, including their distinct mean plate configurations,
oscillation amplitudes, and wake structures. Overall, these results support
the interpretation of the network as a single stiffness-conditioned evolution
operator rather than a collection of independently fitted response curves.

\subsection{Oscillation frequency, phase-space structure, and long rollout}
\label{sec:long-horizon-dynamics}

\begin{figure}[!ht]
    \centering
    \includegraphics[width=0.84\textwidth]
    {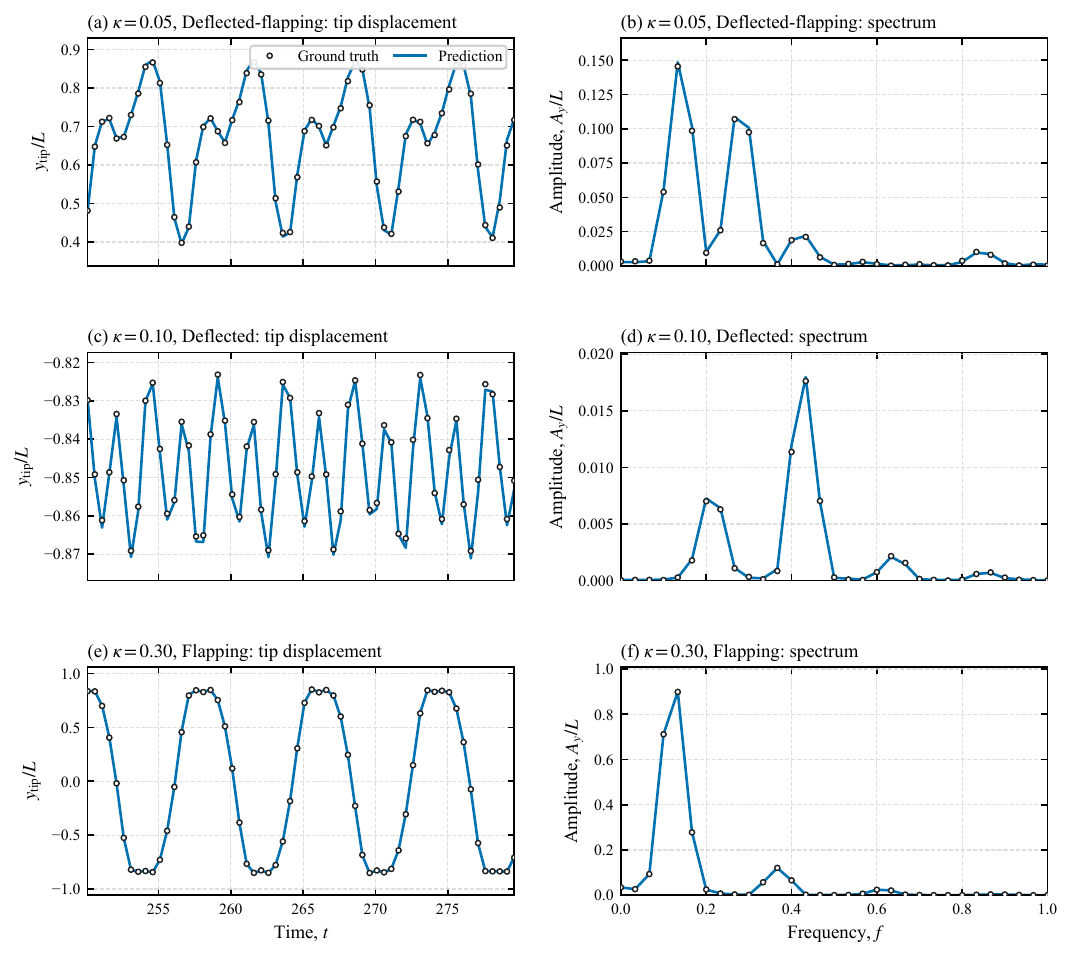}
    \caption{Free-tip displacement histories and frequency spectra for the
    three representative response regimes. The left column compares the
    predicted and CFD free-tip displacements, while the right column shows
    their corresponding amplitude spectra. The rows represent
    (a,b) $\kappa=0.05$, (c,d) $\kappa=0.10$, and
    (e,f) $\kappa=0.30$.}
    \label{fig:tip-displacement-fft}
\end{figure}

Pointwise field errors alone do not establish whether an autoregressive model
preserves the underlying structural dynamics. A small phase drift can increase
the instantaneous field error while leaving the limit-cycle topology and
dominant frequency nearly unchanged. We therefore examine the free-tip
displacement in both the time and frequency domains, as shown in
Fig.~\ref{fig:tip-displacement-fft}.

The dominant frequencies are recovered exactly at the resolution of the
sampled sequence: $f=0.1333$ for $\kappa=0.05$, $f=0.4333$ for
$\kappa=0.10$, and $f=0.1333$ for $\kappa=0.30$. The predicted spectral
peak amplitudes are also close to the CFD values. In particular, the model
distinguishes the higher-frequency, small-amplitude deflected response from
the two lower-frequency flapping responses, rather than reproducing a common
oscillation for all stiffnesses.

\begin{figure}[!ht]
    \centering
    \includegraphics[width=\textwidth]
    {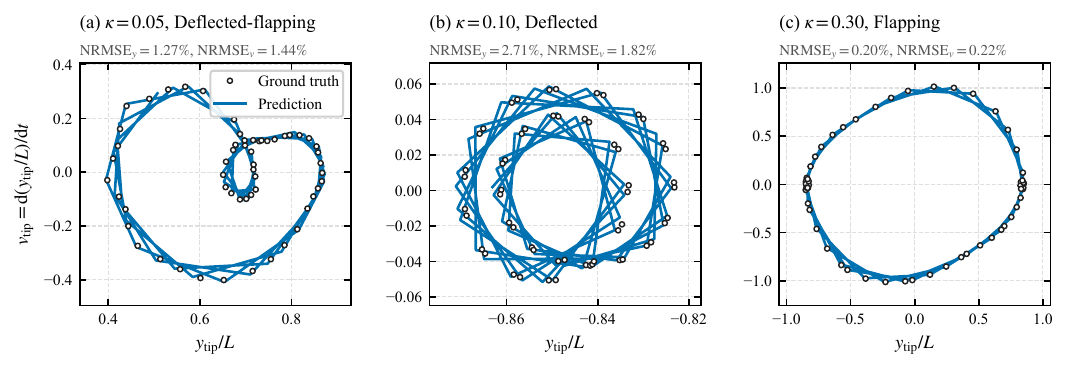}
    \caption{Free-tip phase portraits for three representative response
    regimes: (a) $\kappa=0.05$, deflected--flapping;
    (b) $\kappa=0.10$, deflected; and
    (c) $\kappa=0.30$, flapping. The transverse tip velocity is obtained
    from the time derivative of the free-tip displacement. The normalized
    RMSE values shown above each panel quantify the displacement and velocity
    differences between the prediction and CFD reference.}
    \label{fig:tip-phase-portraits}
\end{figure}

The reconstructed limit cycles are further examined using the free-tip phase
portraits shown in Fig.~\ref{fig:tip-phase-portraits}.

The range-normalized displacement and velocity RMSE values are 1.27\% and
1.44\% for $\kappa=0.05$, 2.71\% and 1.82\% for $\kappa=0.10$, and
0.20\% and 0.22\% for $\kappa=0.30$. The larger normalized displacement
error at $\kappa=0.10$ should be interpreted together with its very small
displacement range: normalization by this range magnifies modest absolute
deviations. The predicted orbit nevertheless retains the location,
orientation, and periodic structure of the CFD cycle.

Finally, a blind autoregressive rollout over 1000 steps was performed to examine long-horizon stability. The predicted coupled state remained bounded and preserved the regime-dependent periodic behavior over the full rollout. Since no synchronized CFD reference is available over this extended interval, this test is interpreted as a stability diagnostic rather than a direct measure of 1000-step prediction accuracy. Detailed stability diagnostics are provided in Appendix A.

\FloatBarrier

\subsection{Interpolation at two unseen stiffnesses}
\label{sec:unseen-stiffness-interpolation}

The 36 training stiffnesses demonstrate that a single stiffness-conditioned
operator can represent the three response regimes over the sampled parameter
interval. To examine interpolation between the discrete training conditions,
two additional CFD trajectories are considered at $\kappa=0.105$ and
$\kappa=0.335$. These stiffnesses were excluded from training and lie midway
between the neighboring training pairs $(0.10,0.11)$ and $(0.33,0.34)$,
respectively. No model parameters are updated for either test. The two cases
represent different interpolation conditions: $\kappa=0.335$ lies within the
established flapping regime, whereas $\kappa=0.105$ lies in the deflected
regime, where symmetry-related plate configurations can occur.


\subsubsection{Interpolation within the flapping regime at
$\kappa=0.335$.}

\begin{figure}[!ht]
    \centering
    \includegraphics[width=0.78\textwidth]
    {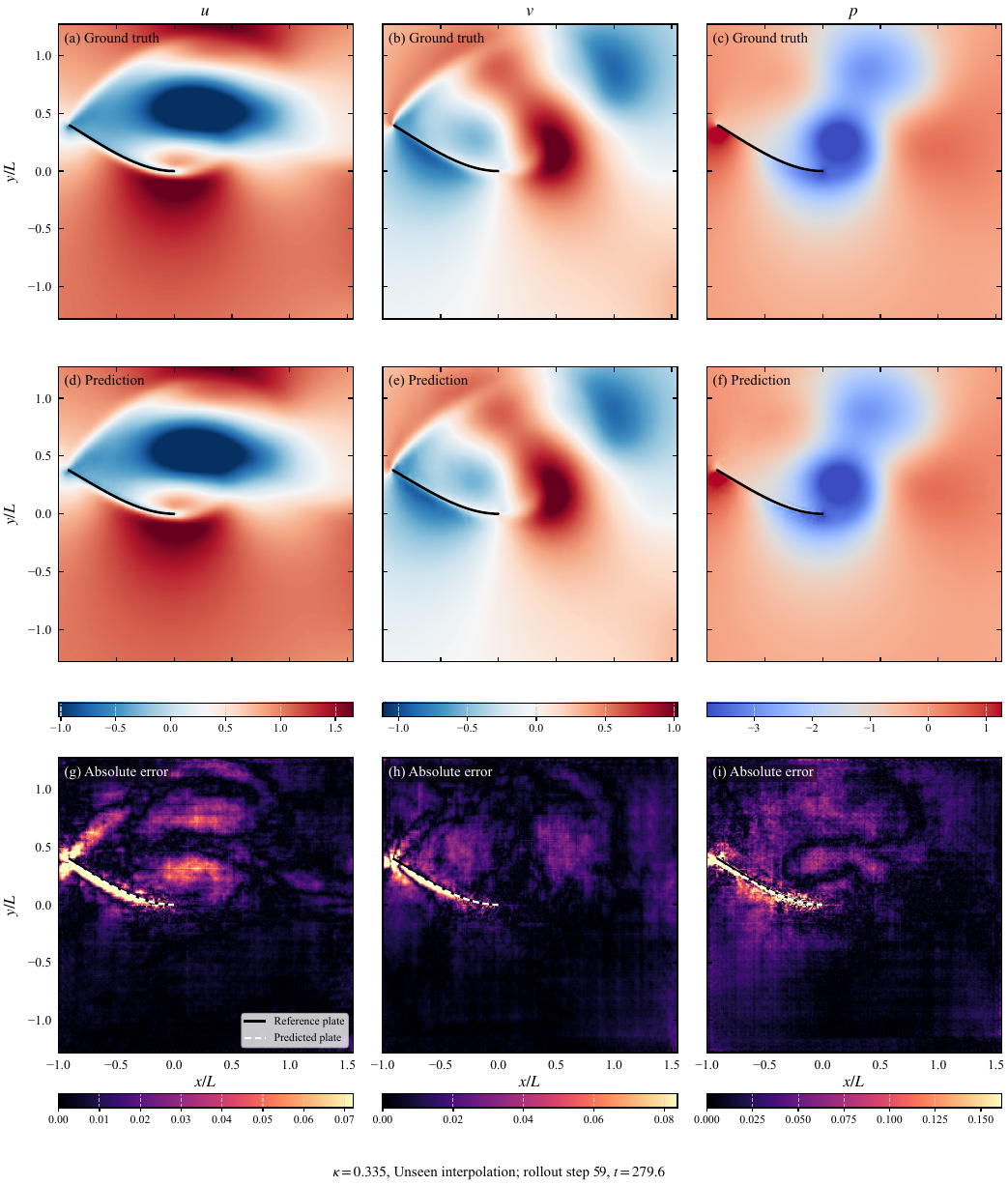}
    \caption{Flow-field reconstruction at rollout step 59 for the unseen
    stiffness $\kappa=0.335$. The columns show $u$, $v$, and $p$, while the
    rows show the CFD ground truth, neural prediction, and pointwise absolute
    error.}
    \label{fig:unseen-0335-fields}
\end{figure}

Figure~\ref{fig:unseen-0335-fields} compares the predicted and CFD flow fields
at the final rollout step.

The model preserves the large-scale flapping pattern, the instantaneous plate
configuration, and the principal spatial organization of the velocity and
pressure fields. Although localized errors remain near the moving plate and
the convected wake structures, the predicted coupled state remains
qualitatively consistent with the CFD solution after 59 recursive steps.

\begin{figure}[!ht]
    \centering
    \includegraphics[width=0.94\textwidth]
    {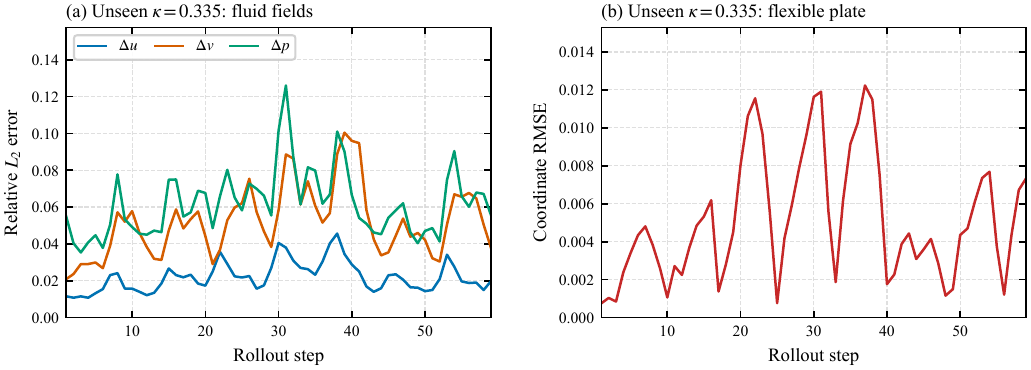}
    \caption{Evolution of the fluid-field relative $L_2$ errors and
    plate-position RMSE over the 59-step autoregressive rollout at the unseen
    stiffness $\kappa=0.335$.}
    \label{fig:unseen-0335-errors}
\end{figure}

The temporal evolution of the prediction errors is shown in
Fig.~\ref{fig:unseen-0335-errors}.

Averaged over the 59 predicted states, the relative errors in $u$, $v$, and
$p$ are 2.20\%, 5.20\%, and 6.19\%, respectively. The mean plate-position
and plate-velocity RMSE values are $5.13\times10^{-3}$ and
$7.29\times10^{-3}$, respectively. At rollout step 59, the corresponding
fluid-field errors are 1.97\%, 3.79\%, and 5.57\%, while the plate-position
and plate-velocity RMSE values are $7.34\times10^{-3}$ and
$8.77\times10^{-3}$. Both the fluid-field and structural errors remain
bounded throughout the evaluated interval, indicating stable interpolation
of the coupled Eulerian and Lagrangian states.

\begin{figure}[!ht]
    \centering
    \includegraphics[width=\textwidth]
    {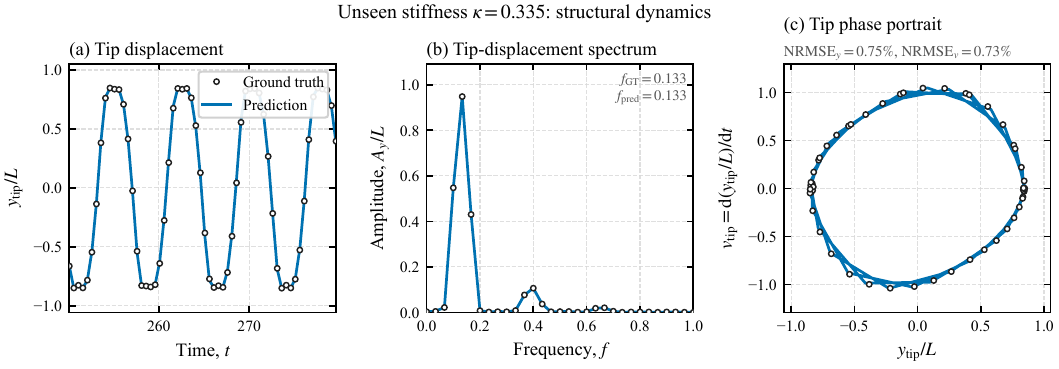}
    \caption{Structural dynamics at the unseen stiffness $\kappa=0.335$:
    free-tip displacement, Fourier spectrum, and phase portrait. CFD values
    are shown by open circles and neural predictions by solid lines.}
    \label{fig:unseen-0335-dynamics}
\end{figure}

The corresponding low-dimensional structural dynamics are presented in
Fig.~\ref{fig:unseen-0335-dynamics}.

The predicted and CFD free-tip signals have the same dominant frequency,
$f=0.1333$. The range-normalized displacement and velocity phase-space RMSE
values are 0.75\% and 0.73\%, respectively. The predicted orbit therefore
retains the amplitude, orientation, and periodic structure of the CFD limit
cycle.

\FloatBarrier

\subsubsection{Interpolation within the deflected regime at
$\kappa=0.105$.}

\begin{figure}[!ht]
    \centering
    \includegraphics[width=0.78\textwidth]
    {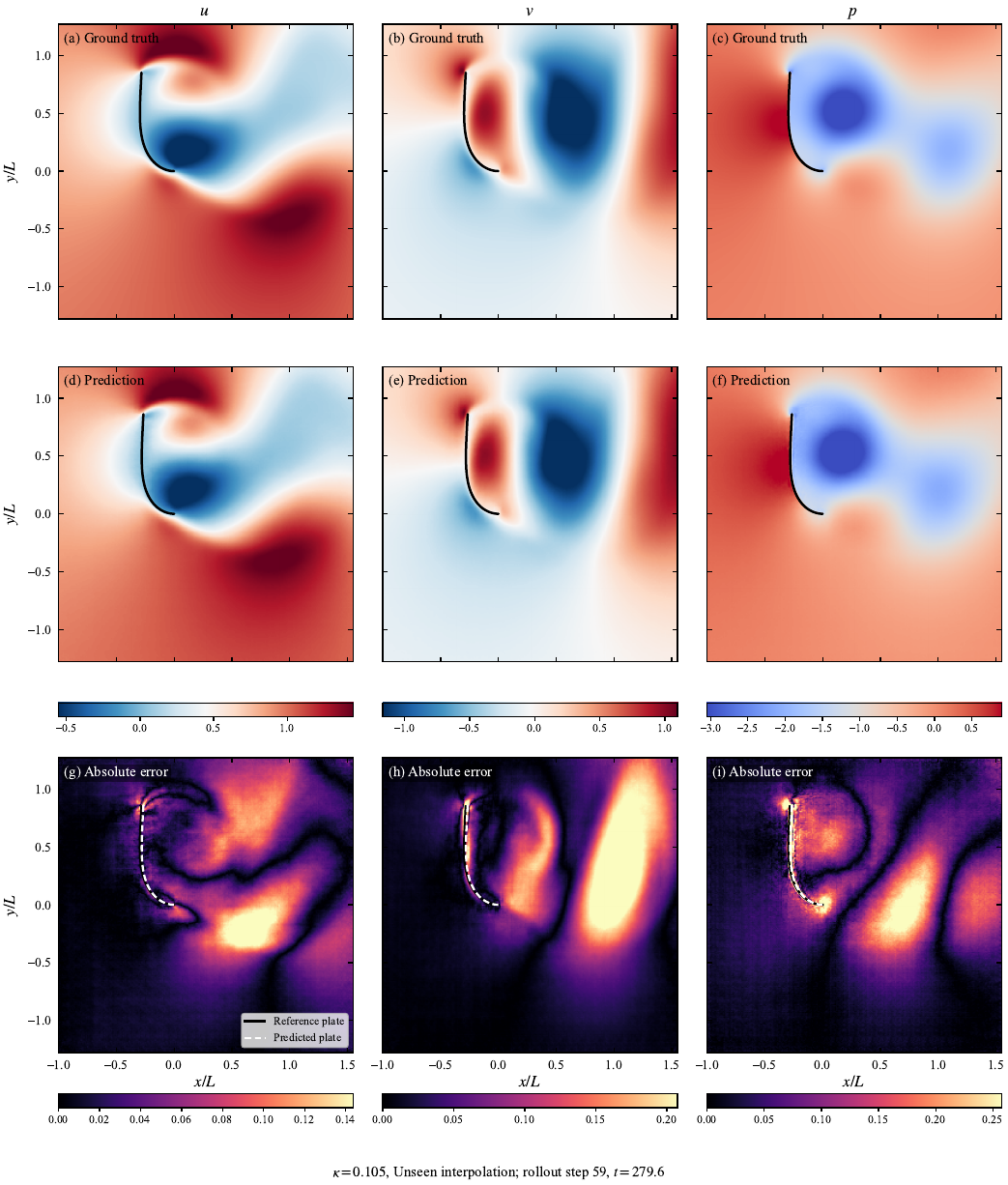}
    \caption{Flow-field reconstruction at rollout step 59 for the unseen
    stiffness $\kappa=0.105$. The columns show $u$, $v$, and $p$, while the
    rows show the CFD ground truth, neural prediction, and pointwise absolute
    error.}
    \label{fig:unseen-0105-fields}
\end{figure}

Interpolation at $\kappa=0.105$ is more sensitive because the plate response
has a small oscillation amplitude and symmetry-related deflected
configurations occur in this stiffness range. Nevertheless, as shown in
Fig.~\ref{fig:unseen-0105-fields}, the model recovers the upward-deflected
mean configuration of the CFD trajectory and retains the principal near-plate
and wake structures.

The pointwise discrepancies are larger than those at $\kappa=0.335$,
particularly in the transverse velocity and pressure fields. The predicted
plate and flow fields nevertheless remain on the same deflected branch as
the CFD reference.

\begin{figure}[!ht]
    \centering
    \includegraphics[width=0.94\textwidth]
    {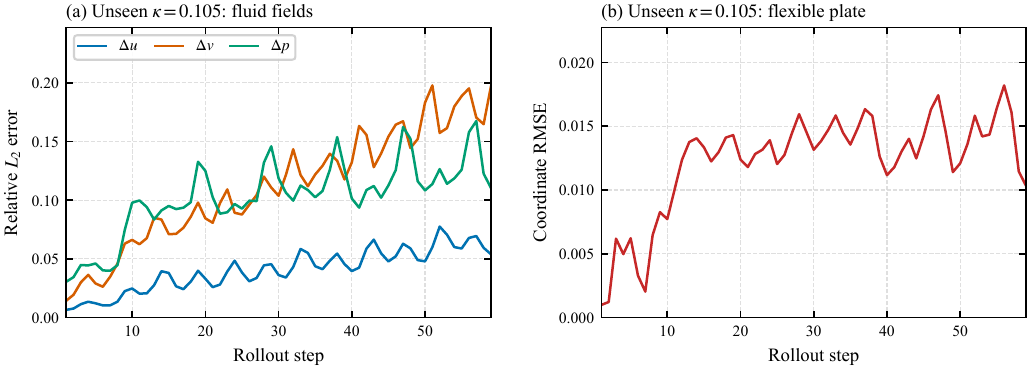}
    \caption{Evolution of the fluid-field relative $L_2$ errors and
    plate-position RMSE over the 59-step autoregressive rollout at the unseen
    stiffness $\kappa=0.105$. The errors remain bounded, although their
    magnitudes are higher than those obtained within the flapping regime.}
    \label{fig:unseen-0105-errors}
\end{figure}

The error histories in Fig.~\ref{fig:unseen-0105-errors} remain bounded,
although their magnitudes are higher than those obtained at
$\kappa=0.335$.

Averaged over the 59 predicted states, the relative errors in $u$, $v$, and
$p$ are 4.05\%, 11.17\%, and 10.28\%, respectively. The mean
plate-position and plate-velocity RMSE values are $1.22\times10^{-2}$ and
$7.98\times10^{-3}$, respectively. At rollout step 59, the corresponding
fluid-field errors are 5.40\%, 19.70\%, and 11.05\%, while the
plate-position and plate-velocity RMSE values are $1.03\times10^{-2}$ and
$5.08\times10^{-3}$. The absence of monotonic error growth indicates that
the increased errors do not arise from numerical divergence of the
autoregressive rollout.

\begin{figure}[!ht]
    \centering
    \includegraphics[width=\textwidth]
    {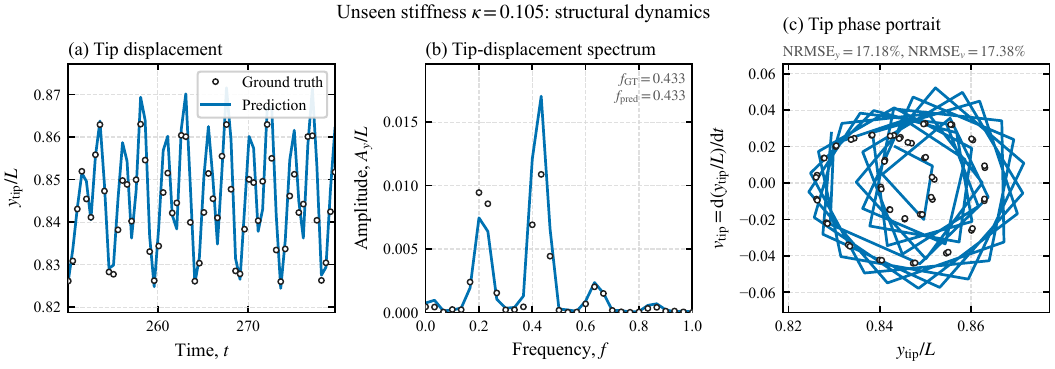}
    \caption{Structural dynamics at the unseen stiffness $\kappa=0.105$:
    free-tip displacement, Fourier spectrum, and phase portrait. Although
    the dominant frequency and mean deflected branch are recovered,
    differences remain in the oscillation amplitude and phase-space
    trajectory.}
    \label{fig:unseen-0105-dynamics}
\end{figure}

The structural response is examined in
Fig.~\ref{fig:unseen-0105-dynamics}.

The model recovers the CFD dominant frequency of $f=0.4333$ and retains a
repeatable oscillatory motion. The range-normalized displacement and velocity
phase-space RMSE values are 17.18\% and 17.38\%, respectively. Thus, the
model captures the mean deflected branch and the characteristic structural
frequency, although differences remain in the oscillation amplitude and
phase. The interpolation at $\kappa=0.105$ is therefore less accurate than
that at $\kappa=0.335$, but it does not exhibit an incorrect branch
orientation or an unstable rollout.

Taken together, the two unseen-stiffness cases show that the same
stiffness-conditioned operator can interpolate both within the flapping
regime and within the branch-sensitive deflected regime. The interpolation
at $\kappa=0.335$ accurately preserves the flow-field organization and
structural limit cycle, while the case at $\kappa=0.105$ retains the correct
mean branch and dominant frequency with larger instantaneous errors. These
results support the use of a single conditional evolution operator across
multiple stiffness-dependent FSI responses, while also identifying reduced
instantaneous accuracy in the small-amplitude deflected regime.

\FloatBarrier

\subsection{Differentiable aerodynamic-force reconstruction from predicted flow fields}
\label{subsec:force_reconstruction}

The preceding results show that the coupled neural operator can maintain a coherent moving interface while evolving the surrounding velocity and pressure fields. Reconstructing the flow and structural states, however, does not by itself provide a complete description of the fluid--structure dynamics. The aerodynamic loading is the quantity through which the fluid drives the structural motion, and the time-dependent drag and lift coefficients therefore provide a stricter physical assessment of the predicted coupled state.

Unlike pointwise flow-field errors, aerodynamic forces are global functionals of the solution. They depend simultaneously on pressure, velocity gradients, vorticity, momentum transport, and their temporal variation. Small local errors that are barely visible in an instantaneous flow-field plot may accumulate coherently during force integration. A predicted field may therefore reproduce the main wake structures while still yielding inaccurate aerodynamic coefficients.

The conventional evaluation of the aerodynamic force is based on the pressure and viscous tractions acting directly on the instantaneous body surface,
\begin{equation}
    \boldsymbol{F}
    =
    \int_{\partial B}
    \left(
        -p\boldsymbol{n}
        +
        \boldsymbol{\tau}\cdot\boldsymbol{n}
    \right)\mathrm{d}s ,
    \label{eq:direct_surface_force}
\end{equation}
where \(\partial B\) denotes the deforming plate boundary. This formulation requires accurate near-wall pressure and velocity gradients. These quantities are particularly sensitive to the spatial resolution of the neural prediction, because the thin boundary layer is only partially resolved on the coarse Eulerian grid. Direct traction integration consequently amplifies near-interface smoothing and local differentiation errors.

To reduce this dependence on the under-resolved near-wall region, \liwei{the} DMT is employed to express the aerodynamic force using quantities evaluated over a control domain and its enclosing contour \cite{wu2005unsteady}. In the present two-dimensional setting, the control surface reduces to a closed control line surrounding the plate. The force evaluation is thereby transferred from the body surface to the better-resolved outer-flow region.

\subsubsection{Fixed control contours for the flexible-plate motion}
\label{subsubsec:fixed_control_contours}

The control contour must enclose the plate throughout the complete autoregressive rollout. A contour that follows the plate at every time step would introduce a time-dependent integration geometry and complicate both the numerical evaluation and subsequent differentiation. A single fixed contour is therefore constructed separately for each stiffness.

Denoting the \(i\)-th predicted Lagrangian point at rollout step \(n\) by \((x_i^n,y_i^n)\), the contour is obtained from the envelope of all predicted plate configurations,
\begin{equation}
\begin{aligned}
    x_0 &= \min_{n,i}(x_i^n)-m, &
    x_1 &= \max_{n,i}(x_i^n)+m,\\
    y_0 &= \min_{n,i}(y_i^n)-m, &
    y_1 &= \max_{n,i}(y_i^n)+m,
\end{aligned}
\qquad m=0.3,
\label{eq:control_contour_bounds}
\end{equation}
after which the four sides are aligned with the Eulerian grid. The margin separates the contour from the immediate vicinity of the plate, while the fixed geometry ensures that the same integration operator is used at every rollout step.

\begin{figure}[!ht]
    \centering
    \includegraphics[width=0.95\linewidth]
    {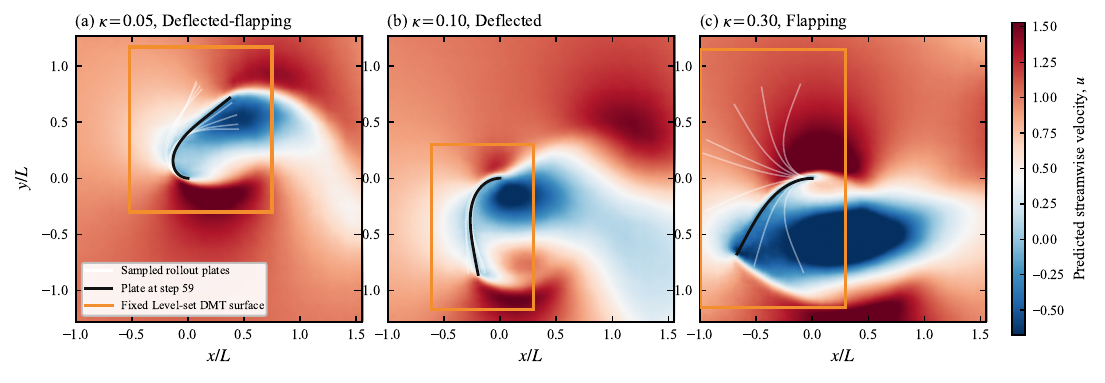}
    \caption{Fixed DMT control contours for the three representative
    flexible-plate regimes. The background shows the predicted streamwise
    velocity at rollout step 59. The white curves show sampled plate
    configurations during the rollout, the black curve denotes the plate at
    step 59, and the orange rectangle is the fixed control contour used for
    force reconstruction.}
    \label{fig:three-regime-control-contours}
\end{figure}

Figure~\ref{fig:three-regime-control-contours} illustrates the control contours for the deflected-flapping, deflected, and flapping regimes. The plate occupies substantially different regions of the domain in the three cases, so a separate contour is required for each stiffness. Within each case, however, the contour remains fixed and encloses the full structural trajectory. This construction separates the force-integration geometry from the instantaneous plate motion and provides a common basis for the explicit and Level-set DMT realizations.

\subsubsection{Isolation of the DMT reconstruction error}
\label{subsubsec:dmt_cfd_validation}

Before applying DMT to the neural-predicted fields, the force operator is evaluated using the original CFD flow fields. This intermediate validation is necessary because a discrepancy in the final force history may originate either from the DMT reconstruction or from the neural flow prediction. Supplying resolved CFD fields removes the second source and isolates the numerical error of the control-contour formulation.

The validation is performed for the flapping case at $\kappa=0.30$ and $Re=200$, using 501 consecutive CFD fields at the original temporal resolution. The pressure-containing explicit Line DMT operator is evaluated on the fixed control contour, and a cubic spline is used to differentiate the moment term in time. The resulting coefficients are compared with the aerodynamic forces reported directly by the CFD solver. The error metrics are computed using the 499 interior samples after excluding the two endpoint samples.

The range-normalized root-mean-square error is defined as
\begin{equation}
    \mathrm{NRMSE}
    =
    \frac{
        \sqrt{
            \frac{1}{N}
            \sum_{n=1}^{N}
            \left(
                \widehat{C}_n-C_n
            \right)^2
        }
    }{
        \max(C)-\min(C)
    }
    \times100\%,
    \label{eq:force_nrmse}
\end{equation}
where \(C\) denotes the direct CFD coefficient and \(\widehat{C}\) its DMT reconstruction. This range-based normalization avoids the singular behavior of pointwise relative errors when the lift coefficient crosses zero.

\begin{figure}[!ht]
    \centering
    \includegraphics[width=0.95\linewidth]
    {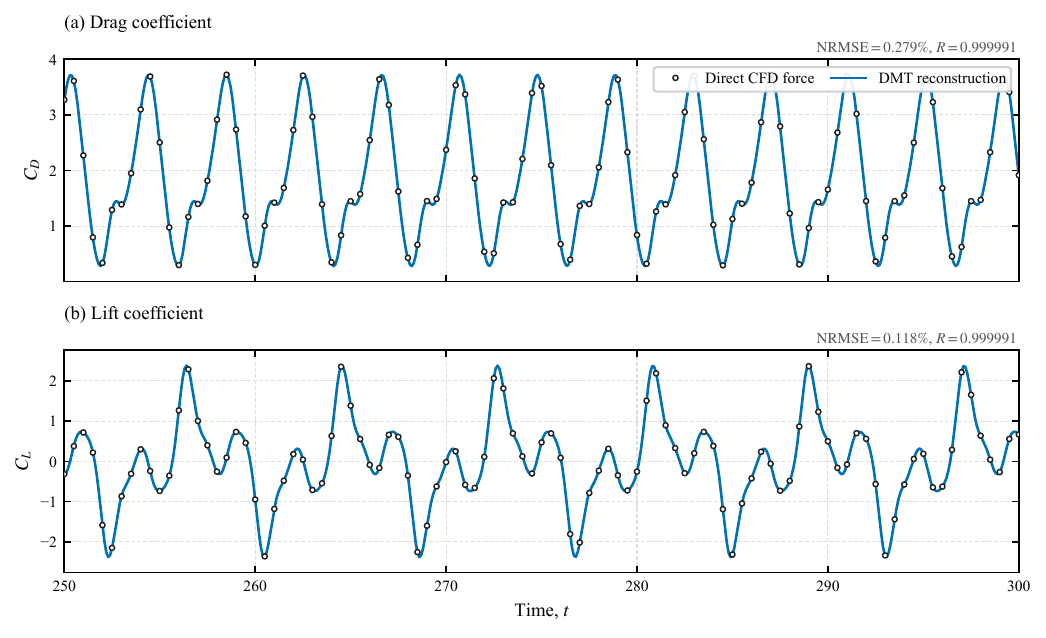}
    \caption{Validation of the pressure-containing explicit Line DMT
    reconstruction using the original CFD flow fields at $\kappa=0.30$.
    Open circles denote the direct CFD force and solid lines denote the DMT
    reconstruction: (a) drag coefficient and (b) lift coefficient.}
    \label{fig:dmt-cfd-validation}
\end{figure}

As shown in Fig.~\ref{fig:dmt-cfd-validation}, the DMT reconstruction is nearly
indistinguishable from the direct CFD force. The drag coefficient gives an
NRMSE of $0.279\%$ and a Pearson correlation coefficient of $0.999991$, while
the lift coefficient gives an NRMSE of $0.118\%$ with the same correlation
coefficient. The agreement extends to both the principal extrema and the
smaller secondary variations of the force histories.

This result establishes that the explicit Line DMT formulation can recover the
aerodynamic loading without direct access to the near-wall traction. More
importantly, it provides a reference level for interpreting the neural-rollout
results: force errors substantially larger than the sub-percent CFD-field
reconstruction error can be attributed primarily to inaccuracies in the
predicted flow fields rather than to the underlying DMT formulation.

\subsubsection{Aerodynamic loading recovered from neural rollouts}
\label{subsubsec:dmt-neural-rollout}

Having isolated the DMT reconstruction error, the same force operator is
applied to the autoregressively predicted $u$, $v$, and $p$ fields. Three
stiffnesses are selected to represent the distinct structural responses
contained in the data set,
\begin{equation}
    \kappa=0.05
    \quad\text{(deflected-flapping)},\qquad
    \kappa=0.10
    \quad\text{(deflected)},\qquad
    \kappa=0.30
    \quad\text{(flapping)}.
    \label{eq:force-representative-kappa}
\end{equation}

For each case, only the initial coupled state is supplied by CFD; all
subsequent fluid and structural states are generated recursively by the neural
operator. The force reconstruction therefore evaluates whether the predicted
sequence retains the global pressure and momentum organization required to
reproduce the aerodynamic loading.

Two realizations are considered. The explicit Line DMT samples and integrates
the required quantities directly along the grid-aligned control contour. The
Level-set DMT represents the same contour implicitly and replaces its line
contribution with a regularized Eulerian-band integral. Both realizations
receive the same neural-predicted flow sequence.

\begin{figure}[!ht]
    \centering
    \includegraphics[width=0.90\linewidth]
    {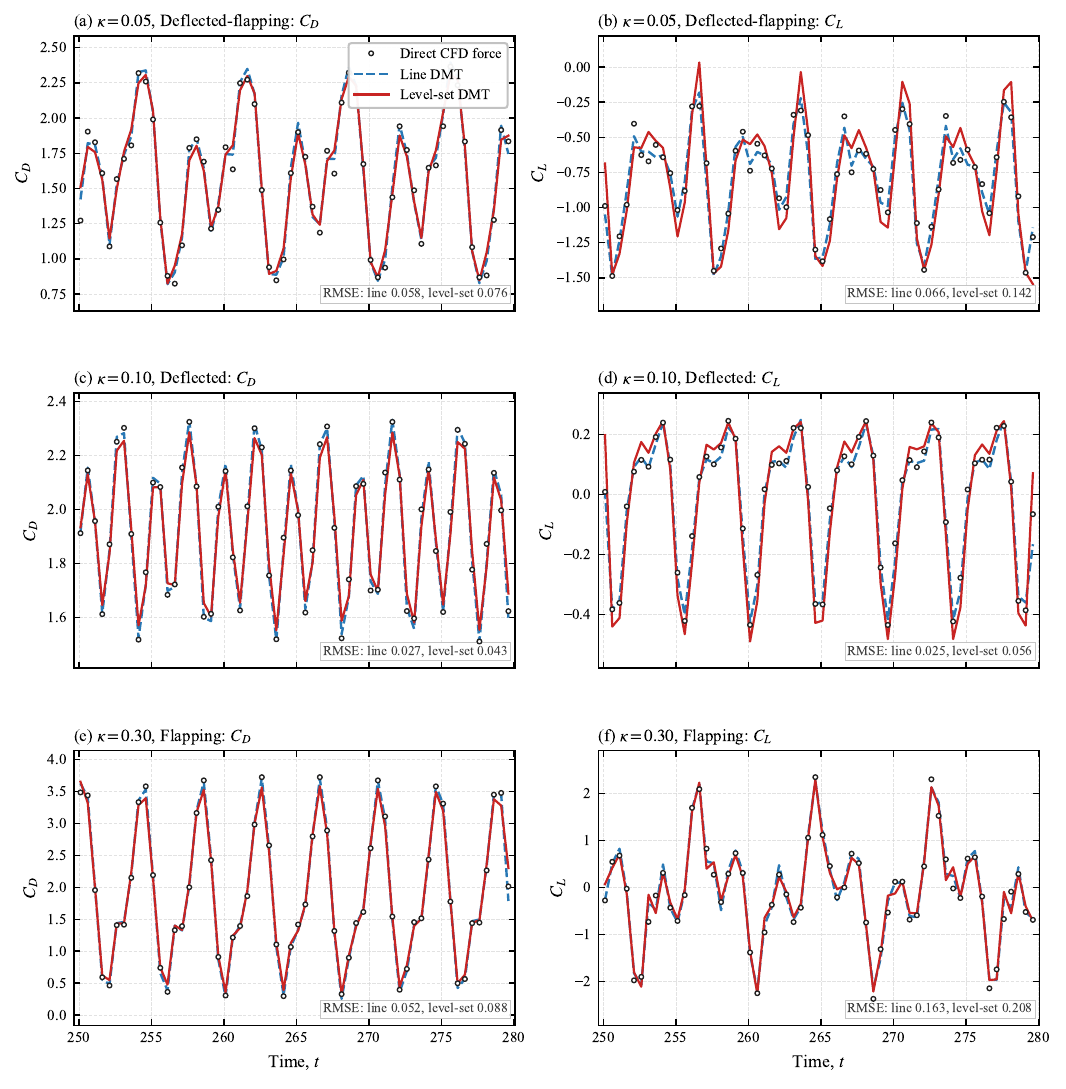}
    \caption{Aerodynamic-force reconstruction from the autoregressively
    predicted flow fields for three representative structural regimes.
    The rows correspond to $\kappa=0.05$ (deflected-flapping),
    $\kappa=0.10$ (deflected), and $\kappa=0.30$ (flapping); the columns
    show $C_D$ and $C_L$. Open circles denote the direct CFD force,
    dashed blue lines denote the explicit Line DMT reconstruction, and
    solid red lines denote the Level-set DMT reconstruction.}
    \label{fig:three-regime-force-comparison}
\end{figure}

Figure~\ref{fig:three-regime-force-comparison} shows that the force histories
remain qualitatively different across the three structural regimes, while
their principal temporal features are consistently recovered from the
neural-predicted fields. The reconstructed curves preserve the dominant peak
locations, oscillation periods, mean loading levels, and phase relationships
of the direct CFD forces.

\begin{table}[!ht]
    \centering
    \small
    \caption{RMSE and range-normalized RMSE (NRMSE) of the aerodynamic
    coefficients reconstructed from the neural-predicted flow fields.}
    \label{tab:three-regime-force-errors}
    \setlength{\tabcolsep}{4.5pt}
    \renewcommand{\arraystretch}{1.15}
    \begin{tabular}{clccccc}
        \hline
        $\kappa$ & Regime & Coefficient &
        \multicolumn{2}{c}{Line DMT} &
        \multicolumn{2}{c}{Level-set DMT}\\
        & & & RMSE & NRMSE (\%) & RMSE & NRMSE (\%)\\
        \hline
        0.05 & Deflected-flapping & $C_D$ & 0.0576 & 3.74 & 0.0765 & 4.96\\
        0.05 & Deflected-flapping & $C_L$ & 0.0663 & 5.34 & 0.1424 & 11.49\\
        0.10 & Deflected          & $C_D$ & 0.0266 & 3.27 & 0.0427 & 5.25\\
        0.10 & Deflected          & $C_L$ & 0.0249 & 3.65 & 0.0558 & 8.19\\
        0.30 & Flapping           & $C_D$ & 0.0521 & 1.52 & 0.0884 & 2.58\\
        0.30 & Flapping           & $C_L$ & 0.1626 & 3.45 & 0.2075 & 4.41\\
        \hline
    \end{tabular}
\end{table}

For the explicit Line DMT, the NRMSE values range from $1.52\%$ to $5.34\%$
across the six aerodynamic-coefficient histories. This agreement indicates
that the flow fields predicted by the stiffness-conditioned operator retain
sufficient outer-flow information to recover the principal aerodynamic
response.

The deflected-flapping case produces a comparatively complex and asymmetric
loading history. Although both DMT realizations recover its major force
variations, the Level-set lift error reaches $11.49\%$, which is the largest
error among the representative cases. The lift signal in this regime contains
several closely spaced extrema and relatively rapid variations, making it
sensitive to accumulated flow-field errors and to the finite-width spatial
averaging introduced by the Level-set band.

For the deflected and flapping cases, the dominant loading periods and phases
remain well preserved. In particular, the two DMT realizations give similar
results for the large-amplitude flapping case at $\kappa=0.30$, despite the
substantial plate excursion. This agreement indicates that the coherent
periodic loading is encoded robustly in the predicted outer-flow dynamics
rather than depending exclusively on the under-resolved near-wall region.

The CFD-field validation in Fig.~\ref{fig:dmt-cfd-validation} establishes a
sub-percent reference error for the explicit Line DMT at $\kappa=0.30$.
The larger Line DMT errors obtained from the neural rollouts therefore
primarily reflect accumulated inaccuracies in the predicted pressure,
velocity, vorticity, and temporal momentum evolution. The additional
difference between the Line and Level-set results also reflects the
finite-width regularization used in the Level-set integration. The force
comparison consequently provides a more integrated physical assessment than
instantaneous field errors because it tests whether the predicted coupled
trajectory remains consistent with the global aerodynamic loading.

These results demonstrate that the neural operator does not merely reproduce
visually similar flow structures. Across three distinct plate-response
regimes, the predicted fields preserve sufficient global information to
reconstruct the dominant drag and lift dynamics. The explicit Line DMT
establishes the feasibility of force recovery from the predicted coarse-grid
fields, while the Level-set realization provides a differentiable
approximation of the same control-contour formulation.

\FloatBarrier

\medskip
\noindent\textit{Aerodynamic loading at the unseen stiffnesses.}

The same force-reconstruction procedure is applied to the two unseen
stiffnesses examined in
Section~\ref{sec:unseen-stiffness-interpolation}. No model parameters are
updated, and both DMT realizations operate on the recursively predicted flow
fields.

At $\kappa=0.335$, the reconstructed forces closely follow the direct CFD
results throughout the evaluated interval.

\begin{figure}[!ht]
    \centering
    \includegraphics[width=0.95\linewidth]
    {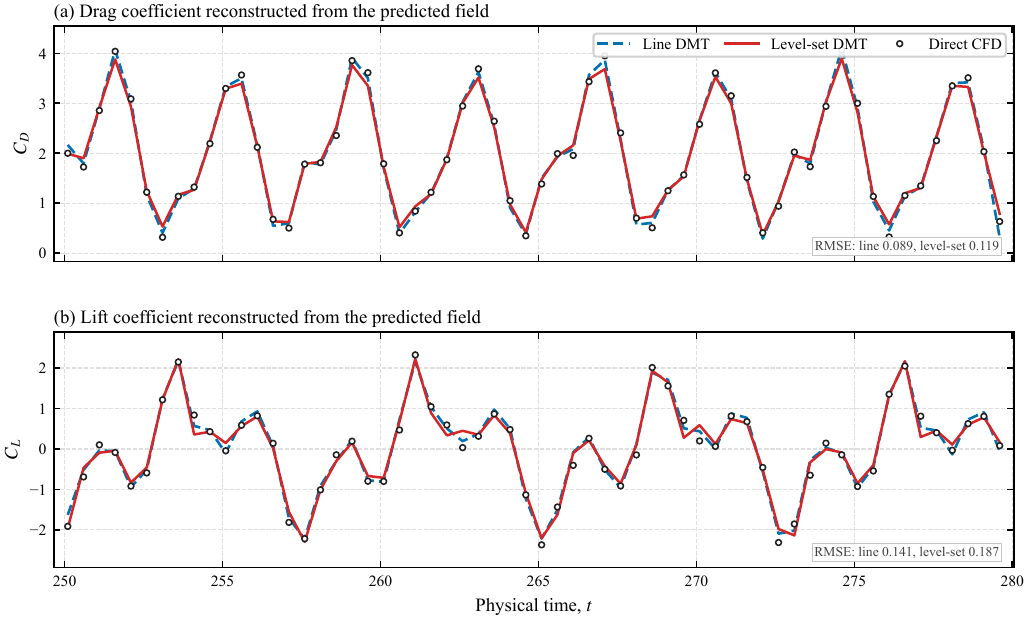}
    \caption{Drag and lift coefficients reconstructed from the predicted
    flow fields at the unseen stiffness $\kappa=0.335$, compared with the
    forces reported directly by the CFD solver. Open circles denote the
    direct CFD force, dashed blue lines denote the explicit Line DMT
    reconstruction, and solid red lines denote the Level-set DMT
    reconstruction.}
    \label{fig:unseen-0335-force-comparison}
\end{figure}

For the explicit Line DMT, the drag and lift NRMSE values are $2.38\%$ and
$3.00\%$, respectively. The corresponding Level-set DMT errors are $3.17\%$
and $3.99\%$. Both realizations retain the principal extrema, oscillation
period, and phase of the direct CFD forces. This agreement confirms that the
interpolated flow fields preserve the global pressure and momentum
organization required to recover the aerodynamic loading within the flapping
regime.

The force reconstruction at $\kappa=0.105$ is more challenging, consistent
with the larger flow-field and structural errors reported in
Section~\ref{sec:unseen-stiffness-interpolation}.

\begin{figure}[!ht]
    \centering
    \includegraphics[width=0.95\linewidth]
    {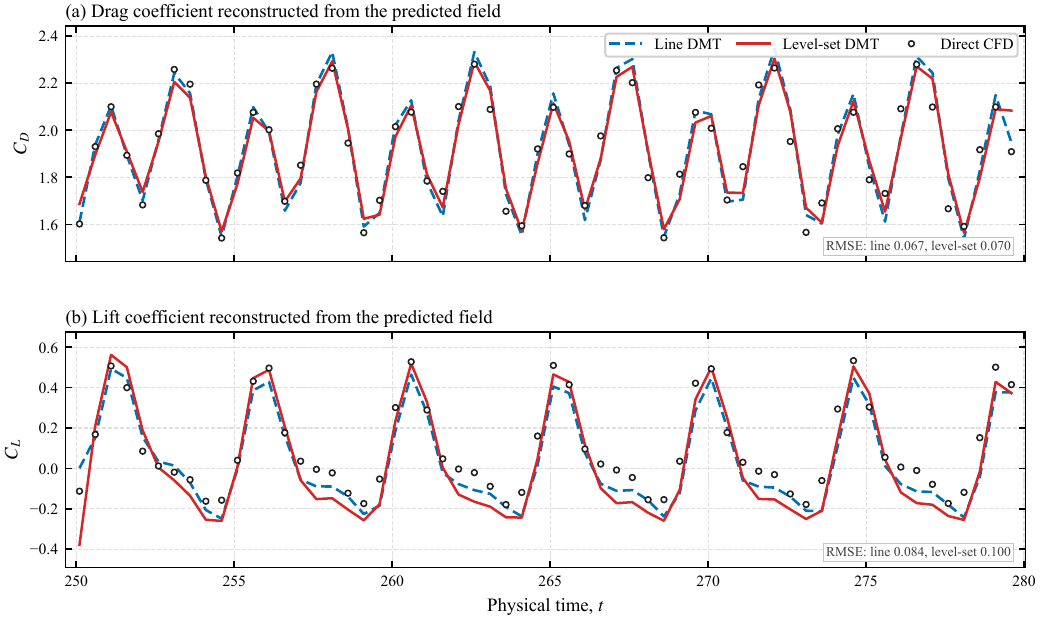}
    \caption{Drag and lift coefficients reconstructed from the predicted
    flow fields at the unseen stiffness $\kappa=0.105$, compared with the
    forces reported directly by the CFD solver. Open circles denote the
    direct CFD force, dashed blue lines denote the explicit Line DMT
    reconstruction, and solid red lines denote the Level-set DMT
    reconstruction.}
    \label{fig:unseen-0105-force-comparison}
\end{figure}

For the explicit Line DMT, the drag and lift NRMSE values are $9.00\%$ and
$11.73\%$, respectively, while the corresponding Level-set DMT values are
$9.50\%$ and $14.07\%$. Although these errors are larger than those at
$\kappa=0.335$, the reconstructed histories retain the principal periodic
variations of the direct CFD forces. 
\liwei{The larger discrepancies are consistent with the higher flow-field and structural errors observed in this branch-sensitive regime, while the bounded rollout errors provide no evidence of autoregressive divergence.}

\FloatBarrier

\subsubsection{Level-set realization and differentiability}
\label{subsubsec:dmt-differentiability}

Explicit control-line integration requires the extraction and ordered
traversal of discrete contour points. Although suitable for post-processing,
such indexing operations are inconvenient within a tensor-based
automatic-differentiation graph. The fixed contour is therefore represented
implicitly as the zero level set of a signed-distance function,
\begin{equation}
    \Sigma
    =
    \left\{
        \boldsymbol{x}\in\Omega:
        \phi(\boldsymbol{x})=0
    \right\}.
    \label{eq:levelset-control-contour}
\end{equation}

A generic contour integral is converted to an Eulerian area integral using a
smoothed Dirac-delta function,
\begin{equation}
    \oint_{\Sigma}
    \mathcal{I}\,\mathrm{d}s
    \approx
    \int_{\Omega}
    \mathcal{I}\,
    \delta_{\varepsilon}(\phi)
    \lvert\nabla\phi\rvert
    \,\mathrm{d}\Omega,
    \label{eq:levelset-line-to-area}
\end{equation}
where $\delta_{\varepsilon}(\phi)$ distributes the contour contribution over
a narrow band of finite width. Because the contour remains fixed, the
signed-distance field, the smoothed delta function, and the normal-vector
weights can be precomputed. The force reconstruction is then composed of
spatial and temporal differences, pointwise tensor products, and weighted
summations, all of which remain in the automatic-differentiation graph. The
resulting discrete operator admits
\begin{equation}
    \frac{\partial(C_D,C_L)}
         {\partial(u,v,p)}
    \label{eq:force-field-gradient}
\end{equation}
for the complete predicted flow sequence.

The differentiable formulation makes the aerodynamic-force reconstruction
available as an active functional of the predicted flow field. A conventional
non-differentiable post-processing procedure can evaluate lift and drag, but
it cannot transmit information on how the upstream flow prediction or a
prescribed design variable should change in order to modify these quantities.
In contrast, the Level-set formulation retains the dependence of the
reconstructed force on every velocity and pressure value participating in
the Eulerian integration.

When the force operator is composed with a differentiable neural rollout, a
force-based objective $\mathcal{J}$ can, in principle, transmit sensitivity
information through the complete computational chain,
\begin{equation}
    \frac{\mathrm{d}\mathcal{J}}{\mathrm{d}\alpha}
    =
    \frac{\partial\mathcal{J}}{\partial\boldsymbol{C}}
    \frac{\partial\boldsymbol{C}}{\partial\boldsymbol{q}}
    \frac{\partial\boldsymbol{q}}{\partial\alpha},
    \qquad
    \boldsymbol{C}=(C_D,C_L),
    \qquad
    \boldsymbol{q}=(u,v,p),
    \label{eq:force-gradient-chain}
\end{equation}
where $\alpha$ denotes a differentiable model parameter, conditioning
variable, control variable, or design variable. The fixed Level-set contour
is advantageous in this setting because its signed-distance field and
quadrature weights remain unchanged during the rollout. Consequently, no
time-dependent contour extraction or discrete boundary-indexing operation
interrupts the gradient path.

The regularized Level-set integration requires a balance between smoothing
and grid resolution. Analytical checks of the contour integral, geometric
gradient, Taylor remainder, and translating-point-vortex force are presented
in ~\ref{app:differentiable_force_verification}. These tests verify
the accuracy of the Level-set approximation and the consistency of its
automatically differentiated gradients.

In the present study, the force operator is applied only as a post-processing
diagnostic, and no force-based objective is included in the training loss.
The differentiable formulation nevertheless provides the computational
connection required for future force-aware learning and gradient-based
aeroelastic design.

Taken together, the results establish a hierarchy of predictive capability.
Across the 36 training stiffnesses, a single stiffness-conditioned operator
reconstructs the coupled flow fields and plate motion over three physically
distinct response regimes, with bounded 59-step prediction errors and stable
long-term autoregressive rollouts. At the unseen stiffness $\kappa=0.335$, the
model accurately interpolates the flow-field organization, structural limit
cycle, and aerodynamic loading within the flapping regime. The more
challenging case at $\kappa=0.105$ exhibits larger instantaneous field,
structural, and force errors, but the prediction recovers the correct
deflected branch and the dominant structural frequency. Together with the
aerodynamic-force results, these findings show that the proposed model
provides a unified representation of stiffness-dependent FSI without
requiring a separate surrogate for each response regime.

\FloatBarrier

\section{Conclusions}
\label{sec:conclusion}
\liwei{
We have developed and systematically evaluated a stiffness-conditioned
neural evolution operator for fluid--structure interaction involving an
inverted flexible plate. The framework jointly evolves Eulerian flow
fields and an explicit Lagrangian structural state, using 101 ordered
structural tokens carrying nodal coordinates and velocities together
with a global stiffness-conditioning token. Bidirectional
fluid--structure coupling enables a single operator to represent the
coupled evolution of the moving plate and surrounding flow.}

\liwei{Across 36 sampled stiffness values, the model reproduced three distinct response regimes, namely deflected--flapping, deflected, and flapping
dynamics, without requiring a separate surrogate for each condition.
The predicted trajectories preserved the principal flow organization,
structural deformation, oscillation frequency, and phase-space
structure. Blind 1000-step autoregressive rollouts remained bounded and
retained the characteristic dynamical behavior of the three regimes,
although these tests assess long-term dynamical stability rather than
pointwise prediction accuracy over the full horizon.}

\liwei{The stiffness-conditioned representation also interpolated to two
unseen stiffness values. The case at $\kappa=0.335$ preserved the
flow-field organization, structural limit cycle, and aerodynamic
response within the flapping regime. At the branch-sensitive
$\kappa=0.105$, the model recovered the same symmetry-related deflected
branch as the CFD reference and retained the dominant oscillatory
behavior, although with larger instantaneous errors.}

\liwei{In addition, a differentiable aerodynamic-force readout was constructed
by combining the derivative-moment transformation with a fixed
control contour represented through a signed-distance function and a
smoothed Dirac-delta formulation. The approach transfers force
evaluation from the under-resolved moving boundary to the better
resolved outer-flow region while retaining gradients with respect to
the predicted flow variables. Validation against resolved CFD fields
confirmed the accuracy of the control-contour formulation, and
application to neural rollouts demonstrated that the principal
aerodynamic loading can be recovered from coarse predicted flow fields.}

\liwei{The differentiable force reconstruction developed here establishes a
direct computational pathway from the learned fluid--structure state to
aerodynamic loading and its gradients. This pathway can be extended
naturally to inverse parameter identification from sparse flow,
structural, or force measurements and to gradient-based optimization of
stiffness and other design variables. The conditional neural operator
also provides a reusable forward model for many-query parameter studies and design iterations, potentially avoiding repeated high-fidelity solutions for every candidate condition. Quantifying this computational
advantage and demonstrating end-to-end inverse and design tasks are
important next steps, together with broader validation under unseen
parameters and incomplete observations.}

\clearpage
\FloatBarrier
\appendix

\renewcommand{\thefigure}{\Alph{section}.\arabic{figure}}
\renewcommand{\theequation}{\Alph{section}.\arabic{equation}}

\section{Long-horizon autoregressive stability diagnostics}
\label{app:long_horizon}

\setcounter{figure}{0}
\setcounter{equation}{0}

To examine the behavior of the learned evolution operator beyond the
59-step evaluation horizon considered in the main text, blind
autoregressive rollouts are performed for 1000 recursive steps at the
three representative stiffnesses,
$\kappa=0.05$, $0.10$, and $0.30$. Only the initial coupled state is
supplied by CFD, and all subsequent fluid and structural states are
generated recursively without intermediate reference-state correction.

\begin{figure}[!t]
    \centering
    \includegraphics[width=0.95\textwidth]
    {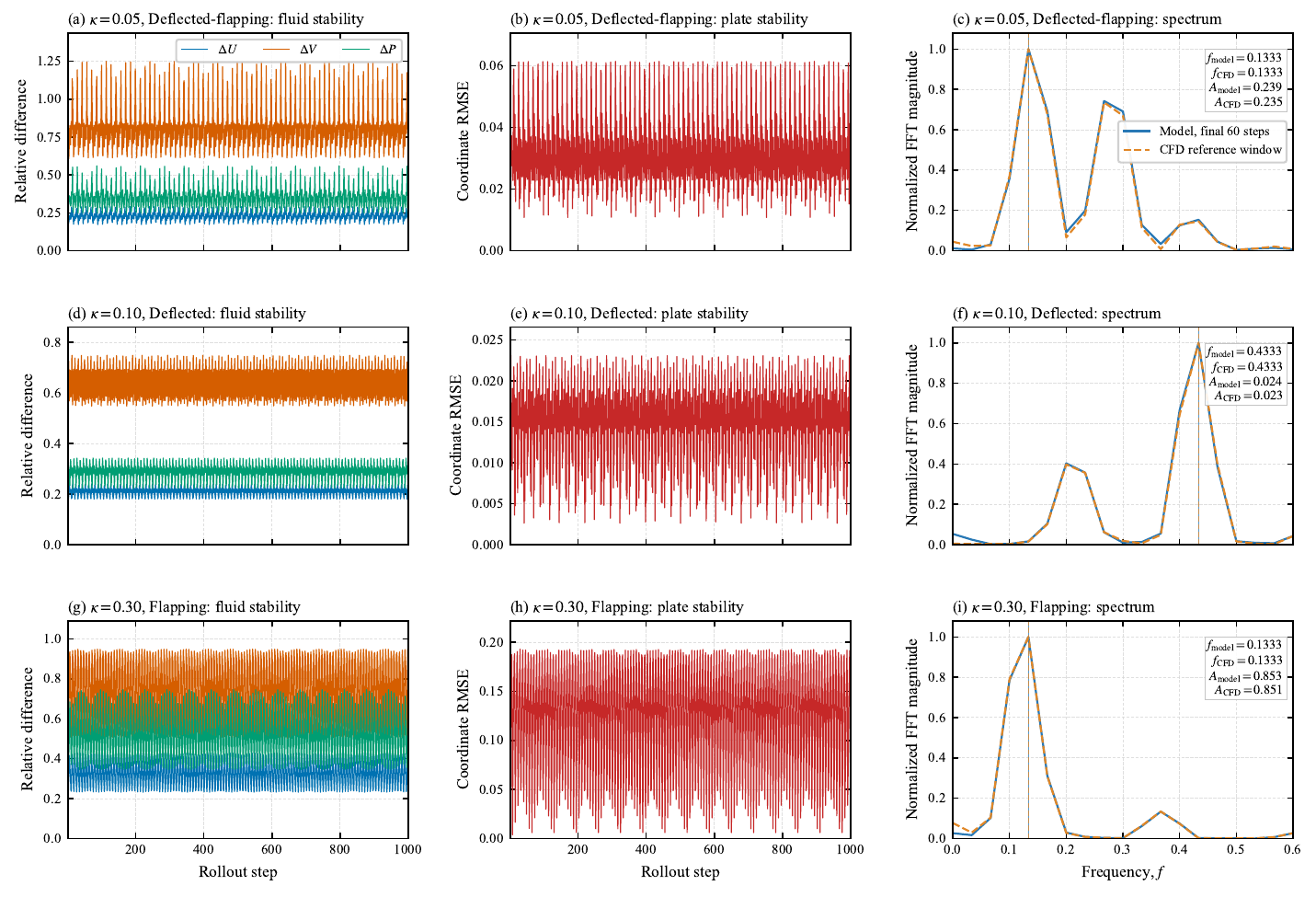}
    \caption{Long-horizon stability diagnostics over 1000 fully
    autoregressive steps for $\kappa=0.05$, $0.10$, and $0.30$.
    The left column shows the relative differences between consecutive
    predicted fluid states, and the middle column shows the coordinate
    RMSE between consecutive predicted plate states. The right column
    compares the normalized spectrum obtained from the final 60
    predicted states with that of the corresponding 60-state CFD
    reference window. These quantities are used as stability
    diagnostics rather than errors against a synchronous 1000-step
    CFD trajectory.}
    \label{fig:long-term-rollout}
\end{figure}

Figure~\ref{fig:long-term-rollout} presents the corresponding
long-horizon stability diagnostics. The relative differences between
consecutive predicted fluid states and the coordinate RMSE between
consecutive predicted plate states are used to identify unbounded
growth, abrupt state changes, or collapse of the autoregressive
trajectory. The free-tip spectra obtained from the final 60 predicted
states are compared with those of the corresponding 60-state,
statistically developed CFD reference windows.

The blind rollouts remain bounded and preserve regime-dependent
periodic behavior over the complete 1000-step interval. The spectra
obtained from the final 60 predicted states retain the dominant
frequencies of the corresponding CFD reference trajectories. Repeated
reuse of the predicted coupled state does not lead to immediate
divergence, collapse to a steady solution, or unphysical growth of the
plate motion.

Because a synchronized CFD reference trajectory is not available over
the complete extended interval, these results should be interpreted as
evidence of long-horizon numerical and dynamical stability rather than
as a direct measurement of 1000-step prediction accuracy.

\FloatBarrier

\section{Analytical verification of the differentiable force reconstruction}
\label{app:differentiable_force_verification}

\setcounter{figure}{0}
\setcounter{equation}{0}

\begin{figure}[!htbp]
    \centering
    \includegraphics[width=\linewidth]
    {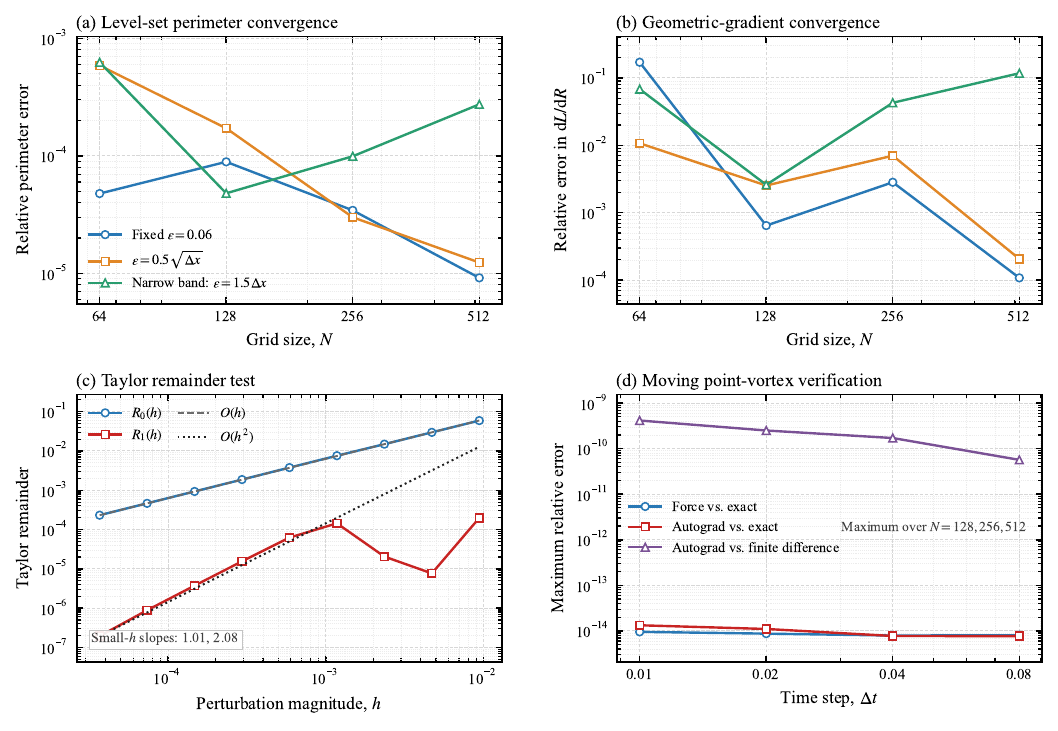}

    \caption{Analytical verification of the Level-set
    integration and discrete gradients:
    (a) circular-perimeter convergence;
    (b) geometric-gradient convergence;
    (c) Taylor-remainder test; and
    (d) translating-point-vortex force and gradient
    verification.}
    \label{fig:differentiability_validation}
\end{figure}

The Level-set quadrature and its geometric gradient are
first tested on a circular contour with exact perimeter
$L=2\pi R$ and derivative $dL/dR=2\pi$.
Fig.~\ref{fig:differentiability_validation}(a,b) compares
the corresponding errors under grid refinement and
different smoothing-band choices.
Both quantities approach their analytical values when
the regularized band is adequately resolved.
Gradient consistency is further assessed using
\begin{equation}
\begin{aligned}
R_0(h) &= |J(R+h)-J(R)|,\\
R_1(h) &= |J(R+h)-J(R)-hJ'(R)|.
\end{aligned}
\label{eq:app_force_taylor}
\end{equation}
The fitted small-perturbation slopes are 1.01 and 2.08,
consistent with the expected first- and second-order
behavior, respectively
(Fig.~\ref{fig:differentiability_validation}(c)).

The vorticity-moment force is additionally tested using
a point vortex of strength $\Gamma$ translating with
velocity $(c_x,c_y)$, for which
\begin{equation}
\mathbf{F}=\Gamma(-c_y,c_x),
\qquad
\frac{\partial\mathbf{F}}{\partial\Gamma}=(-c_y,c_x).
\label{eq:app_force_point_vortex}
\end{equation}
The computed force and its automatic-differentiation
gradient agree with the analytical expressions to
approximately machine precision.
Differences between automatic differentiation and
centered finite differences are of order
$10^{-11}$--$10^{-10}$
(Fig.~\ref{fig:differentiability_validation}(d)).

\clearpage

\section*{Acknowledgments}
This work was supported by the National Natural Science Foundation of China under Grant Nos. 92470120 and 12672385.

\bibliographystyle{elsarticle-num}
\bibliography{refs}

\end{document}